\documentclass[11pt]{article}
\usepackage[dvips]{graphicx}
\usepackage{bm}
\usepackage{amsmath}
\usepackage{amssymb}
\usepackage{float}
\usepackage{url}
\renewcommand{\baselinestretch}{1.1}
\renewcommand{\thepage}{}
\makeatletter
\@addtoreset{equation}{section}
\renewcommand{\theequation}{\thesection.\@arabic\c@equation}
\makeatother
\renewcommand{\thefootnote}{\fnsymbol{footnote}}
\begin{document}
\begin{titlepage}
\title{
\vspace*{-4ex}
\hfill{}\\
\hfill
\begin{minipage}{3.5cm}
\end{minipage}\\
\bf On the spectrum around numerical solutions\\
in Siegel gauge in open string field theory
\vspace{3ex}
}

\author{
Isao~{\sc Kishimoto}\footnote{ikishimo@rs.socu.ac.jp}
\\
\vspace{0.5ex}\\
\\
{\it Center for Liberal Arts and Sciences, Sanyo-Onoda City University,}\\
{\it Daigakudori 1-1-1, Sanyo-Onoda Yamaguchi 756-0884, Japan}
\vspace{2ex}
}

\date{}
\maketitle

%
\vspace{7ex}

\begin{abstract}
\normalsize
In bosonic open string field theory, the spectrum around the numerical tachyon vacuum solution in Siegel gauge was investigated by Giusto and Imbimbo. Using their numerical method, we study the mass spectrum around two other solutions, which are ``double brane'' and ``single brane" solutions in Siegel gauge constructed by the level truncation approximation. The ``double brane'' solution was constructed by Kudrna and Schnabl and its energy might correspond to a double brane. On the other hand, the ``single brane'' solution was constructed by Takahashi and the author in the theory around the identity-based solution for the tachyon vacuum and its energy corresponds to the perturbative vacuum, namely, a single brane. From the eigenvalues of the matrix for the kinetic term in Siegel gauge, we find a tachyon state and a massless vector state in the ghost number one sector around the ``single brane'' solution, which is consistent with the perturbative vacuum, although the mass spectrum around the ``double brane'' solution is obscure up to the truncation level ten and within scalar and vector states.

\end{abstract}

\vspace{5ex}

\end{titlepage}

\renewcommand{\thepage}{\arabic{page}}
\renewcommand{\thefootnote}{\arabic{footnote}}
\setcounter{page}{1}
\setcounter{footnote}{0}
%
\renewcommand{\baselinestretch}{1}
\tableofcontents
\renewcommand{\baselinestretch}{1.1}

\section{Introduction and summary
\label{sec:Introduction}}

String field theory is a candidate of a nonperturbative formulation of string theory and we expect that it can describe various physical phenomena.
The action of bosonic open string field theory is given by
\begin{align}
&S[\Psi]
=-\frac{1}{g^2}\left(\frac{1}{2}\langle\Psi,Q_{\rm B}\Psi\rangle+\frac{1}{3}\langle\Psi,\Psi\ast\Psi\rangle\right)
\label{eq:actionQB}
\end{align}
and its equation of motion is
\begin{align}
&Q_{\rm B}\Psi+\Psi\ast\Psi=0.
\label{eq:EOMQB}
\end{align}
As a numerical solution to this, the tachyon vacuum solution in Siegel gauge, which we denote as $\Psi_{\rm T}$, was constructed by the level truncation approximation, and its energy was evaluated by Sen and Zwiebach \cite{Sen:1999nx}. 
It was shown that it is consistent with the interpretation that $\Psi_{\rm T}$ corresponds to the tachyon vacuum where a D-brane vanishes.
Namely, for the normalized energy 
\begin{align}
E[\Psi]=1-2\pi^2 g^2S[\Psi]
\label{eq:energyE}
\end{align}
which corresponds to the number of D-branes, $E[\Psi_{\rm T}]$  almost vanishes.
Such a computation was performed for higher truncation level $L$ \cite{Moeller:2000xv, Gaiotto:2002wy, Kishimoto:2011zza, Kudrna:2018mxa}.
As a consistency check on $E[\Psi_{\rm T}]\simeq 0$, the gauge-invariant observable\footnote{
This is also called the gauge-invariant overlap \cite{Kawano:2008ry} or the Ellwood invariant \cite{Ellwood:2008jh} in the literature.
} $E_0[\Psi]=1-2\pi \langle I|V|\Psi\rangle$
with the identity string field $|I\rangle$ and an on-shell closed string vertex operator $V$,
which is expected to coincide with $E[\Psi]$ for a class of solutions to the equation of motion \cite{Baba:2012cs}, was evaluated and confirmed that $E_0[\Psi_{\rm T}]\simeq 0$. From Ref.~\cite{Kudrna:2018mxa}, more precise values are
$E[\Psi_{\rm T}]=-0.000627118$ and 
$E_0[\Psi_{\rm T}]=0.0120671$ at the truncation level $L=30$.

The theory around the tachyon vacuum solution $\Psi_{\rm T}$ was explored in Refs.~\cite{Hata:2001rd, Ellwood:2001py, Ellwood:2001ig}.
In particular, the spectrum of the theory around $\Psi_{\rm T}$ was investigated numerically by Giusto and Imbimbo \cite{Giusto:2003wc, Imbimbo:2006tz} up to the truncation level $L=10$, and then it was found that there are nontrivial cohomologies at the ghost number $g=-1,0,3$, and $4$, where we regard the ghost number of the conformal vacuum $|0\rangle$ as zero, and it is compatible with the interpretation as the tachyon vacuum,
although it is different from a naive expectation that there is no cohomology due to no brane.

In this paper, we apply the numerical method in Refs.~\cite{Giusto:2003wc, Imbimbo:2006tz} to two other solutions in Siegel gauge, which we denote as $\Psi_{\rm D}$ and $\Phi_{\rm S}$, to study the spectrum of the theory around them.
We restrict the states to scalar and vector for simplicity and perform computations up to the truncation level $L=10$ in the same way as Ref.~\cite{Imbimbo:2006tz} for the theory around $\Psi_{\rm T}$.

$\Psi_{\rm D}$ is the ``double brane" solution constructed by Kudrna and Schnabl in Ref.~\cite{Kudrna:2018mxa} numerically, which might correspond to a double brane because the energy of $\Psi_{\rm D}$
was evaluated as $E[\Psi_{\rm D}]\simeq 2$ up to the truncation level $L=28$.
More precisely, according to Ref.~\cite{Kudrna:2018mxa}, 
it is $E[\Psi_{\rm D}]=1.8832-0.161337i$ ($L=28$),
although the gauge invariant observable $E_0[\Psi_{\rm D}]=1.32953+0.178426i$ ($L=28$), which is closer to $1$ than $2$.
The imaginary parts of $E[\Psi_{\rm D}]$ and $E_0[\Psi_{\rm D}]$ arise from that of $\Psi_{\rm D}$ itself.
Namely, $\Psi_{\rm D}$ does not satisfy the reality condition at least up to $L=28$,
 although there is a possibility that it becomes real in the large-$L$ limit.
 Thus we do not know whether or not $\Psi_{\rm D}$ can be interpreted as a double brane solution literally.
Numerical solutions  obtained by kinds of continuous deformations of $\Psi_{\rm D}$ were constructed in Refs.~\cite{Kishimoto:2020vfg, Kishimoto:2021ubd} and evaluated their energy and gauge-invariant observable, but the physical interpretation of $\Psi_{\rm D}$ remained unclear.
Instead of the values of $E[\Psi_{\rm D}]$ and $E_0[\Psi_{\rm D}]$,  we here investigate the mass spectrum around $\Psi_{\rm D}$ to extract its physical meaning.
However, we could not find meaningful numerical trends in our results.

$\Phi_{\rm S}$ was constructed by Takahashi and the author in Ref.~\cite{Kishimoto:2009nd} numerically. It is a solution to the equation of motion in the theory around the identity-based solution $\Psi_{\rm TT}$ corresponding to the tachyon vacuum, which is the scalar solution constructed by Takahashi and Tanimoto in Ref.~\cite{Takahashi:2002ez} with $a=-1/2$ where $a$ is a real parameter included in the solution. This action is given by
\begin{align}
S_{\Psi_{\rm TT}}[\Phi]\equiv S[\Psi_{\rm TT}+\Phi]-S[\Psi_{\rm TT}]=-\dfrac{1}{g^2}\left(\frac{1}{2}\langle\Phi,Q_{\rm TT}\Phi\rangle+\frac{1}{3}\langle\Phi,\Phi\ast\Phi\rangle\right),
\label{eq:S_PsiTT}
\end{align}
and hence $\Phi_{\rm S}$ is a solution to its equation of motion
\begin{align}
&Q_{\rm TT}\Phi+\Phi\ast\Phi=0.
\label{eq:EOMTT}
\end{align}
We note that $\Psi_{\rm TT}+\Phi_{\rm S}$ is  a solution to the equation of motion in Eq.~(\ref{eq:EOMQB}) and $\Psi_{\rm TT}+\Phi_{\rm S}\ne 0$.
The theory of $S_{\Psi_{\rm TT}}[\Phi]$ has no cohomology in the ghost number $g=1$ sector \cite{Kishimoto:2002xi}.
Although it is difficult to evaluate the energy $E[\Psi_{\rm TT}]$ and the gauge-invariant observable $E_0[\Psi_{\rm TT}]$ directly due to singular properties of the identity string field in $\Psi_{\rm TT}$, some indirect evidences of $E[\Psi_{\rm TT}]=0$ and $E_0[\Psi_{\rm TT}]=0$ were found by numerical calculations \cite{Takahashi:2003ppa, Kishimoto:2009nd}, and later they were shown by some analytical methods \cite{Zeze:2014qha, Kishimoto:2014lua, Ishibashi:2014mua}. 
Therefore, $\Psi_{\rm TT}$, which is not in Siegel gauge, has been expected to represent the tachyon vacuum.
Based on the theory with Eq.~(\ref{eq:S_PsiTT}), we define the normalized energy and gauge-invariant observable as
\begin{align}
&E^{\prime}[\Phi]=-2\pi^2 g^2S_{\Psi_{\rm TT}}[\Phi],
&&E_0^{\prime}[\Phi]=-2\pi \langle I|V|\Phi\rangle,
\end{align}
where these are the same as $E[\Psi_{\rm TT}+\Phi]$ and $E_0[\Psi_{\rm TT}+\Phi]$, respectively,  under $E[\Psi_{\rm TT}]=0$ and $E_0[\Psi_{\rm TT}]=0$.
Then, the numerical solution $\Phi_{\rm S}$ in Siegel gauge gives
$E^{\prime}[\Phi_{\rm S}]\simeq 1$ and $E_0^{\prime}[\Phi_{\rm S}]\simeq 1$
(more precisely, $E^{\prime}[\Phi_{\rm S}]=1.1571287$ and  $E_0^{\prime}[\Phi_{\rm S}]=0.914281$ at the truncation level $L=26$ \cite{Kishimoto:2011zza}),
and hence we can expect that it represents a single brane or the perturbative vacuum.
Here, we investigate the mass spectrum around $\Phi_{\rm S}$ to confirm this interpretation.
We have found that our results are consistent with the expectation. Concretely, we have found one scalar state with $\alpha^{\prime}m^2\simeq -1$ and one vector state with $\alpha^{\prime}m^2\simeq 0$ both in the ghost number $g=1$ sector, which correspond to the tachyon state and the massless vector state on a single D-brane.

This paper is organized as follows.
In Sect.~\ref{sec:Method} we explain our strategy to find the mass spectrum around numerical solutions.
In Sect.~\ref{sec:Dresult} we show the results for the ``double brane" solution $\Psi_{\rm D}$.
In Sect.~\ref{sec:Sresult} we comment on our calculation with $Q_{\rm TT}$ instead of $Q_{\rm B}$ and show the results for the ``single brane" solution $\Phi_{\rm S}$.
In Sect.~\ref{sec:conclusion} we present some concluding remarks on our results.
In Appendix~\ref{sec:Tresults} we review some results for the tachyon vacuum solution $\Psi_{\rm T}$ for comparison.

\section{Our strategy for the mass spectrum around numerical solutions
\label{sec:Method}}

We describe our strategy to find the mass spectrum around numerical solutions using the numerical method in Ref.~\cite{Imbimbo:2006tz}.
Expanding the action of bosonic cubic open string theory in Eq.~(\ref{eq:actionQB})
around a solution $\varPsi_0$ to the equation of motion in Eq.~(\ref{eq:EOMQB}),
we have
\begin{align}
S_{\varPsi_0}[\Phi]\equiv S[\varPsi_0+\Phi]-S[\varPsi_0]=-\dfrac{1}{g^2}\left(\frac{1}{2}\langle\Phi,Q^{\prime}\Phi\rangle+\frac{1}{3}\langle\Phi,\Phi\ast\Phi\rangle\right)
\label{eq:S_varPsi0}
\end{align}
where $Q^{\prime}$ is the BRST operator around the solution $\varPsi_0$ and it is defined by
\begin{align}
Q^{\prime}\phi&=Q_{\rm B}\phi+\varPsi_0\ast\phi-(-1)^{|\phi|}\phi\ast\varPsi_0.
\label{eq:Qprimedef}
\end{align}
Here, $(-1)^{|\phi|}$ is $+1$ ($-1$) if $\phi$ is Grassmann even (odd).
Imposing the Siegel gauge condition $b_0\Phi=0$ on the action $S_{\varPsi_0}[\Phi]$ (\ref{eq:S_varPsi0}), the kinetic term is evaluated by $L_0^{\prime}\equiv\{b_0,Q^{\prime}\}$.
We note that we should include states with all ghost numbers in $\Phi$ after the gauge fixing.
In the following, we restrict $\varPsi_0$ to a solution with zero momentum, and we fix a basis of states
$\{|e^{(g)}_r(p)\rangle\}$ with ghost number $g$ and momentum $p^{\mu}$.
Then, we define a matrix $(L_0^{\prime(g)}(p))_{rs}$ by
\begin{align}
&\langle e_{r_1}^{(2-g)}(p_1),c_0L_0^{\prime}e_{r_2}^{(g)}(p_2)\rangle
=(G^{(g)}L_0^{\prime(g)}(p_2))_{r_1r_2}(2\pi)^{26}\delta^{26}(p_1+p_2),
\end{align}
where $\langle ~,~\rangle$ is the BPZ inner product and $(G^{(g)})_{rs}$ is given by a normalization of the basis:
\begin{align}
\langle e_{r}^{(2-g)}(p_1),c_0e_{s}^{(g)}(p_2)\rangle
=(G^{(g)})_{rs}(2\pi)^{26}\delta^{26}(p_1+p_2).
\end{align}
By expanding the determinant of the matrix $(L_0^{\prime(g)}(p))_{rs}$ as
\begin{align}
&\det L_0^{\prime(g)}(p) =a_g(p^2+m^2)^{d_g}\left(1+O(p^2+m^2)\right),&&
(a_g\ne 0)
\end{align}
we can read off the number of states with a mass $m$ from $d_g$.
However, it is difficult to evaluate the determinant directly, and hence we investigate the numerical behavior of the eigenvalues of the matrix $(G^{(g)}L_0^{\prime(g)}(p))_{rs}$ as a function of $p^2$, especially focusing on those with the smallest absolute value.

To perform numerical calculations, we take a basis with the ghost number $g$ in Siegel gauge of the form
\begin{align}
&a_{l_1}^{\mu_1\dagger}a_{l_2}^{\mu_2\dagger}\cdots
b_{-n_1}\cdots b_{-n_k}c_{-m_1}\cdots c_{-m_{g+k-1}}c_1|p\rangle
\label{eq:basisstate}
\end{align}
($l_1\ge l_2\ge \cdots \ge 1$, $n_1>\cdots >n_k\ge 1$, $m_1>\cdots>m_{g+k-1}\ge 1$)
with a momentum $p^{\mu}=(p^0,0,\cdots,0)$,
where $a_l^{\mu\dagger}=\alpha_{-l}^{\mu}/\sqrt{l}$ is a normalized mode of the matter $\partial X^{\mu}$.
Following the level truncation method,
we restrict the level of states $l_1+l_2+\cdots +n_1+\cdots +n_k+m_1+\cdots +m_{g+k-1}$ up to $L$.
 
 Furthermore, because we only consider the theory around twist-even Lorentz invariant solutions with zero momentum, the state space can be divided into some sectors.
 We divide the level truncated states into two sectors: twist-even and twist-odd, where levels of states are even and odd respectively.
 As for the Lorentz indices, we investigate two sectors: scalar and vector of ${\rm SO}(25)$, which is the little group for  $p^{\mu}=(p^0,0,\cdots,0)$.
 The scalar consists of states in Eq.~(\ref{eq:basisstate}) whose indices $\mu=i=1,2,\dots,25$ are contracted, and the vector consists of states with one uncontracted index $\mu=i$.
 Therefore, we investigate eigenvalues of $(G^{(g)}L_0^{\prime(g)}(p))_{rs}$ for four sectors:
 scalar twist-even, scalar twist-odd, vector twist-even, and vector twist-odd. 
In the tachyonic region $p^2>0$,  we take $p^{\mu}=(0,0,\cdots,p^{25})$ and consider scalar and vector sectors with respect to SO(1,24), which are obtained by replacement $a^{0\dagger}_l$ with $a^{25\dagger}_l$ and taking the index $i$ as  $i=0,1,\dots,24$ in the above.

The dimension of the level truncated space of states up to $L$ in Siegel gauge for scalar (vector) twist-even/odd sector in the ghost number $g$ with a fixed momentum, which we denote as $N^{{\rm s}(g)}_{L({\rm e}/{\rm o})}$  ($N^{{\rm v}(g)}_{L({\rm e}/{\rm o})}$), is given in Table~\ref{tab:Nscalar} (Table~\ref{tab:Nvector}). We note that the relations:
\begin{align}
&N_{L({\rm e})}^{{\rm s}(g)}
=N_{L({\rm e})}^{{\rm s}(2-g)},
&&N_{L({\rm o})}^{{\rm s}(g)}
=N_{L({\rm o})}^{{\rm s}(2-g)},
&&N_{L({\rm e})}^{{\rm v}(g)}
=N_{L({\rm e})}^{{\rm v}(2-g)},
&&N_{L({\rm o})}^{{\rm v}(g)}
=N_{L({\rm o})}^{{\rm v}(2-g)}.
\end{align}

\begin{table}[htbp]
\begin{minipage}[c]{0.5\hsize}
\centering
          \caption{
           Number of scalar states $N^{{\rm s}(g)}_{L({\rm e}/{\rm o})}$
              \label{tab:Nscalar}}
\begin{tabular}{l|r|r|r|r|r}
\hline
$L$&$g=1$&$g=2$&$g=3$&$g=4$&$g=5$\\
\hline\hline
 0(even)&1&0&0&0&0\\
1(odd)&1&1&0&0&0\\
2(even)&5&2&0&0&0\\
3(odd)&9&6&1&0&0\\
4(even)&24&13&2&0&0\\
5(odd)&45&30&7&0&0\\
6(even)&99&61&14&1&0\\
7(odd)&183&125&35&2&0\\
8(even)&363&240&68&7&0\\
9(odd)&655&458&145&15&0\\
10(even)&1216&841&272&36&1\\
\hline
      \end{tabular}
\end{minipage}
\begin{minipage}[c]{0.5\hsize}
\centering
          \caption{
          Number of vector states $N^{{\rm v}(g)}_{L({\rm e}/{\rm o})}$
             \label{tab:Nvector}}
\begin{tabular}{l|r|r|r|r}
\hline
$L$&$g=1$&$g=2$&$g=3$&$g=4$\\
\hline\hline
 0(even)&0&0&0&0\\
1(odd)&1&0&0&0\\
2(even)&2&1&0&0\\
3(odd)&7&3&0&0\\
4(even)&16&9&1&0\\
5(odd)&40&22&3&0\\
6(even)&85&52&10&0\\
7(odd)&184&113&24&1\\
8(even)&367&238&59&3\\
9(odd)&730&478&127&10\\
10(even)&1385&936&272&25\\
\hline
      \end{tabular}
\end{minipage}
\end{table}
          
We fix a basis of the twist-even/odd sector in the ghost number $g$ with momentum $p^{\mu}$
and denote it as $\{e^{(g)}_{r({\rm e}/{\rm o})}(p)\}$.
Using this, we define a matrix $C^{(g)}_{rs({\rm e}/{\rm o})}(p)$, which corresponds to $(G^{(g)}L_0^{\prime(g)}(p))_{rs}$, as
\begin{align}
\langle e_{r({\rm e}/{\rm o})}^{(2-g)}(-p),c_0L_0^{\prime}e_{s({\rm e}/{\rm o})}^{(g)}(p)\rangle&
=C^{(g)}_{rs({\rm e}/{\rm o})}(p)(2\pi)^{26}\delta^{26}(0)
\label{eq:Cgpmmat_def}
\end{align}
and then it satisfies a relation
\begin{align}
C^{(g)}_{rs({\rm e}/{\rm o})}(p)&=(-1)^{1-g}C^{(2-g)}_{sr({\rm e}/{\rm o})}(-p).
\label{eq:Cgpm_sym}
\end{align}
In our calculation of the matrix $C^{(g)}_{rs({\rm e}/{\rm o})}(p)$ for the truncation level $L$, we use twist-even numerical solutions $\varPsi_0$ in Eq.~(\ref{eq:Qprimedef}), which have been obtained by the $(\tilde{L},3\tilde{L})$ level truncation.
In the case that $L$ is even (odd), we take $\tilde{L}=L$ ($\tilde{L}=L+1$).

\section{Numerical results for the ``double brane'' solution
\label{sec:Dresult}}

Similarly to the case of the tachyon vacuum solution $\Psi_{\rm T}$ in Refs.~\cite{Giusto:2003wc, Imbimbo:2006tz}, which we review in Appendix \ref{sec:Tresults},  we show our numerical results for the ``double brane" solution $\Psi_{\rm D}$ in this section.

\begin{figure}[htbp]
\includegraphics[width=8.4cm]{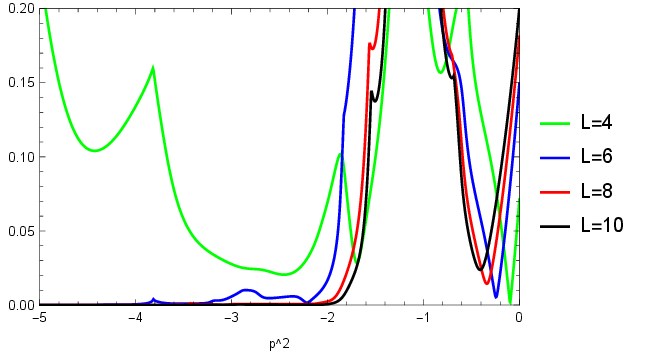}
\hfill
\includegraphics[width=8.4cm]{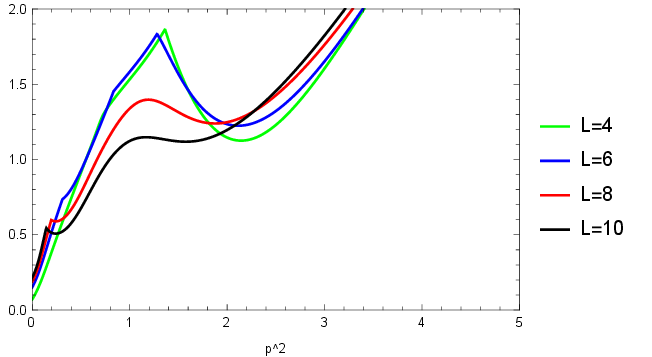}\\

\includegraphics[width=8.4cm]{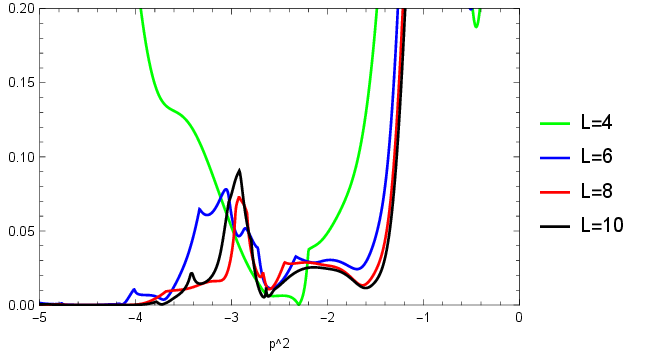}
\hfill
\includegraphics[width=8.4cm]{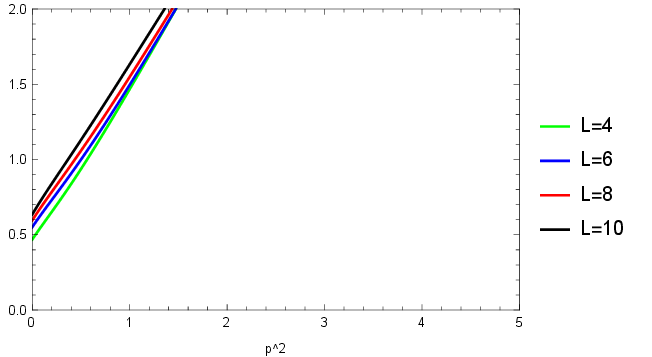}\\

\includegraphics[width=8.4cm]{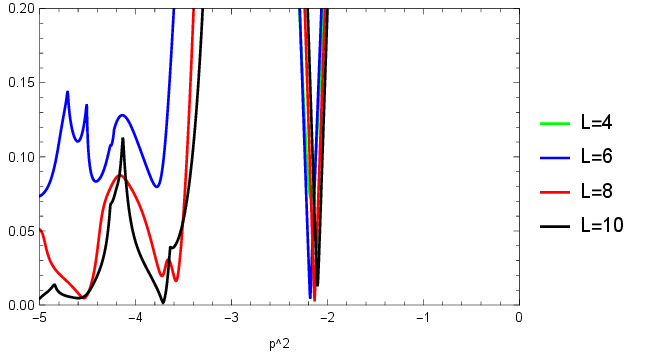}
\hfill
\includegraphics[width=8.4cm]{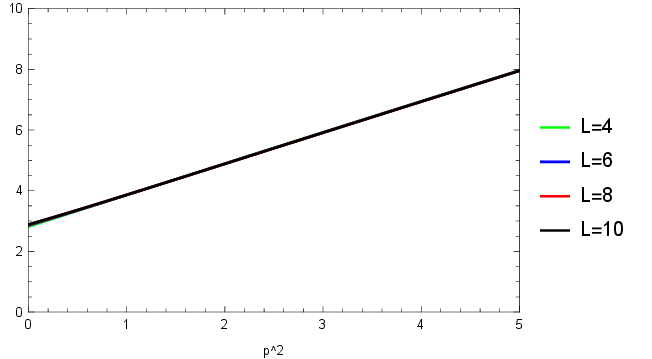}
\caption{
The smallest absolute value of eigenvalues of $C^{(g)}_{rs({\rm e})}(p)$ for $\Psi_{\rm D}$ against $\alpha^{\prime}p^2$ in the scalar twist-even sector for the ghost number $g=1$ (upper), $g=2$ (middle), and $g=3$ (lower).
\label{fig:Dsolscalareven_Abs}
}
\end{figure}

\begin{figure}[htbp]
\includegraphics[width=8.4cm]{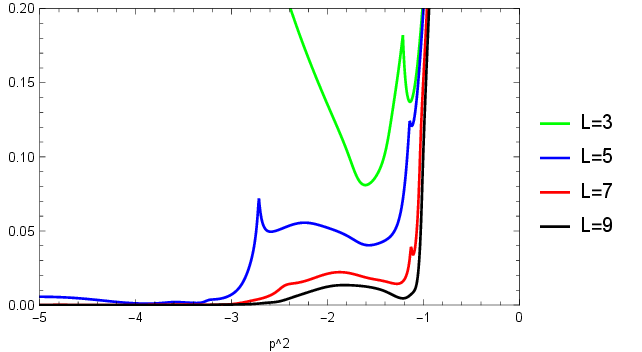}
\hfill
\includegraphics[width=8.4cm]{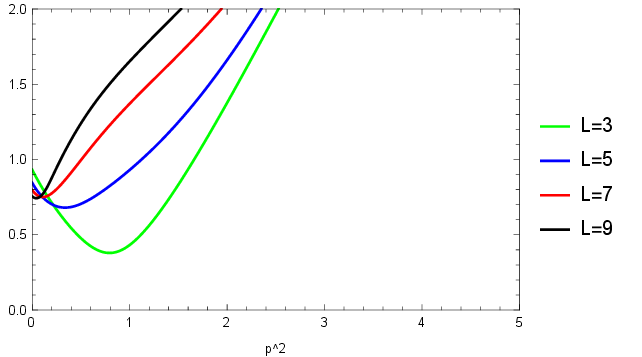}\\

\includegraphics[width=8.4cm]{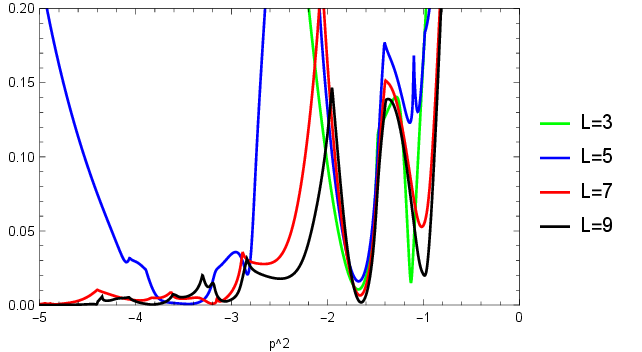}
\hfill
\includegraphics[width=8.4cm]{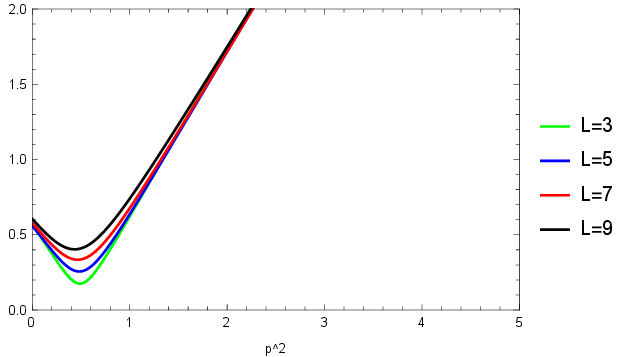}\\

\includegraphics[width=8.4cm]{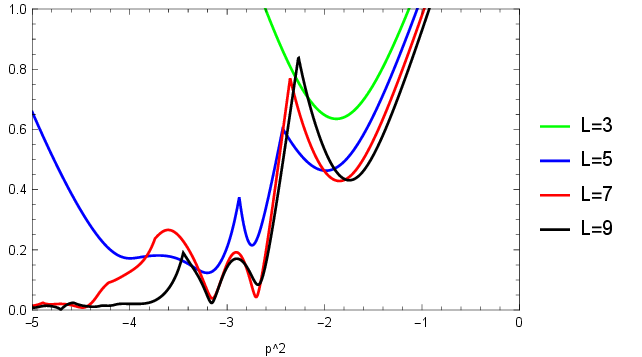}
\hfill
\includegraphics[width=8.4cm]{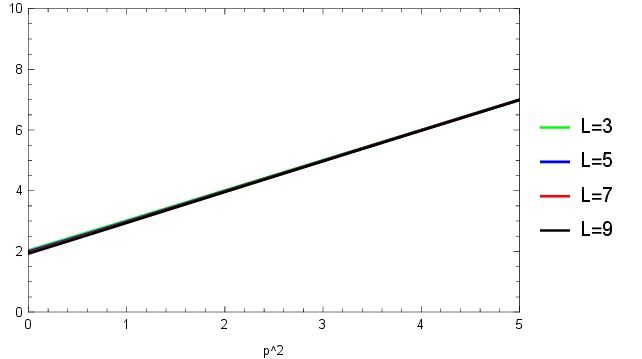}
\caption{
The smallest absolute value of eigenvalues of $C^{(g)}_{rs({\rm o})}(p)$ for $\Psi_{\rm D}$ against $\alpha^{\prime}p^2$ in the scalar twist-odd sector for the ghost number $g=1$ (upper), $g=2$ (middle), and $g=3$ (lower).
\label{fig:Dsolscalarodd_Abs}
}
\end{figure}

\begin{figure}[htbp]
\includegraphics[width=8.4cm]{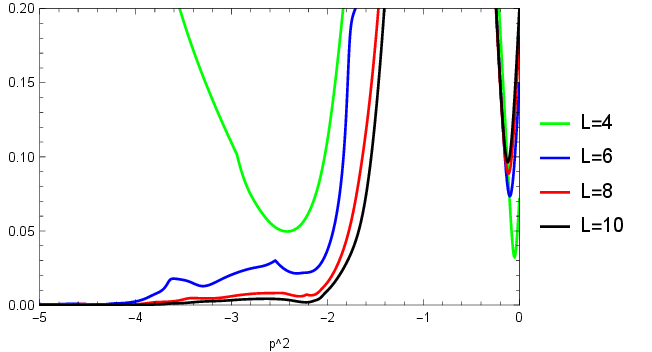}
\hfill
\includegraphics[width=8.4cm]{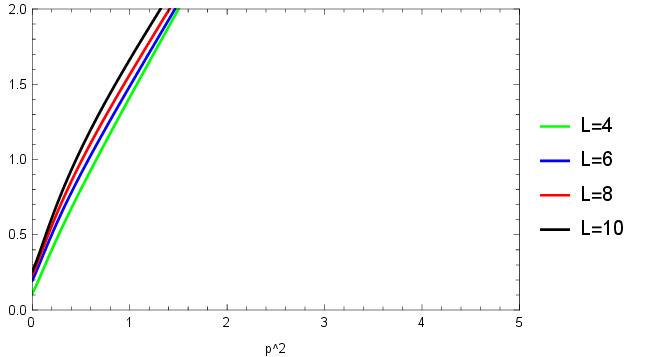}\\

\includegraphics[width=8.4cm]{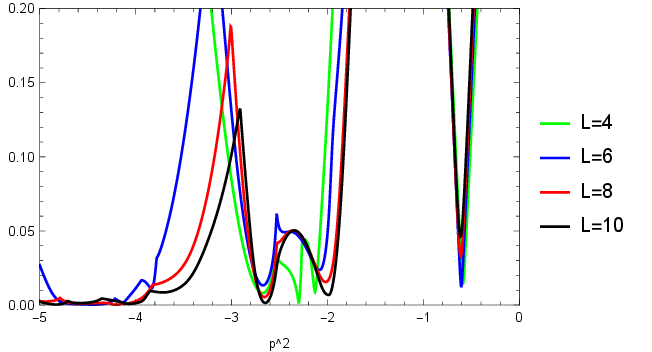}
\hfill
\includegraphics[width=8.4cm]{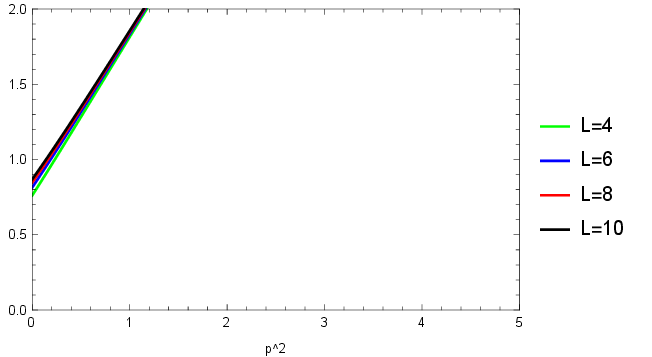}\\

\includegraphics[width=8.4cm]{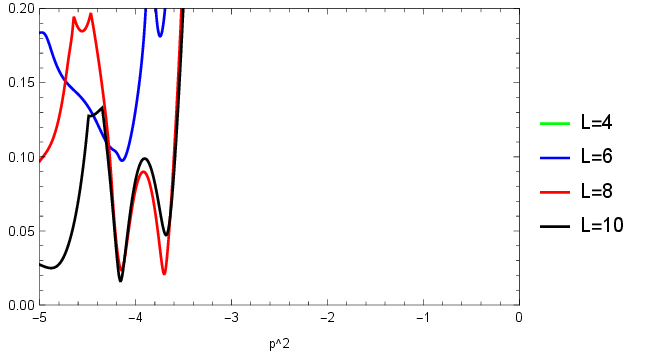}
\hfill
\includegraphics[width=8.4cm]{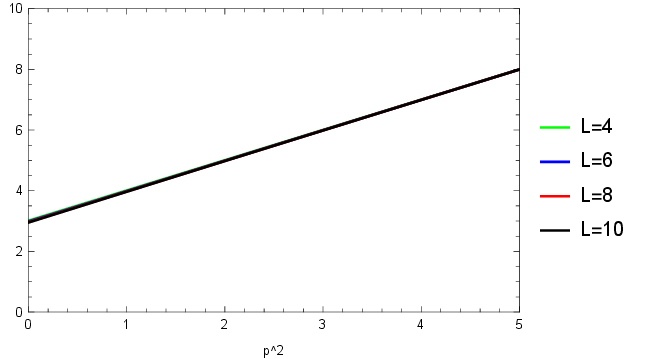}
\caption{
The smallest absolute value of eigenvalues of $C^{(g)}_{rs({\rm e})}(p)$ for $\Psi_{\rm D}$ against $\alpha^{\prime}p^2$ in the vector twist-even sector for the ghost number $g=1$ (upper), $g=2$ (middle), and $g=3$ (lower).
\label{fig:Dsolvectoreven_Abs}
}
\end{figure}

\begin{figure}[htbp]
\includegraphics[width=8.4cm]{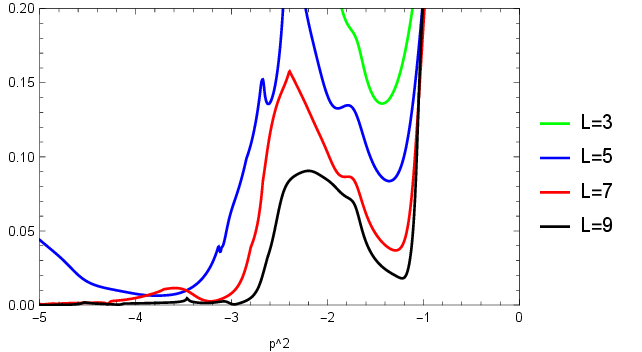}
\hfill
\includegraphics[width=8.4cm]{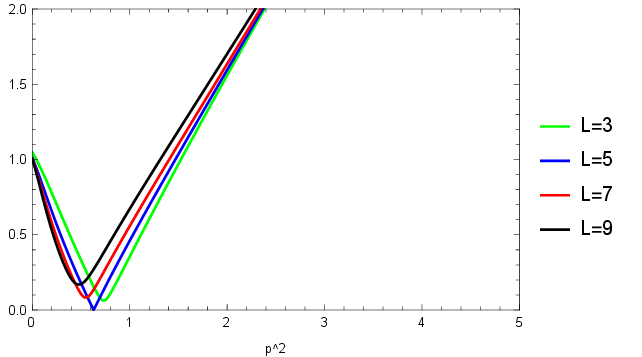}\\

\includegraphics[width=8.4cm]{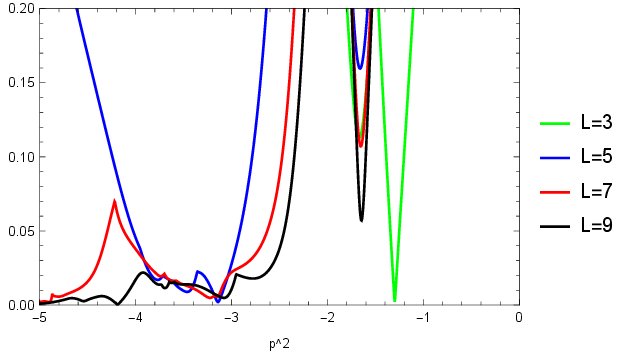}
\hfill
\includegraphics[width=8.4cm]{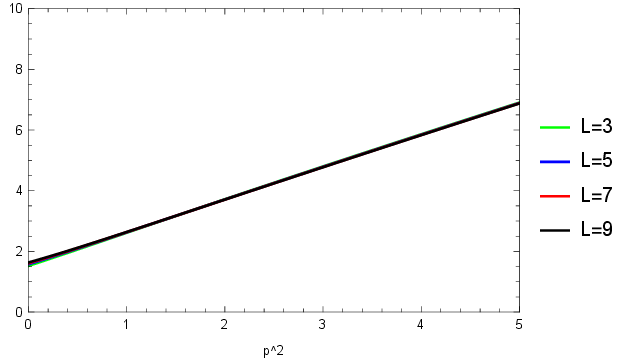}\\

\includegraphics[width=8.4cm]{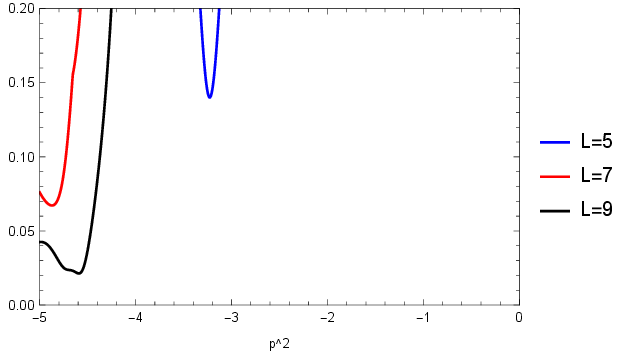}
\hfill
\includegraphics[width=8.4cm]{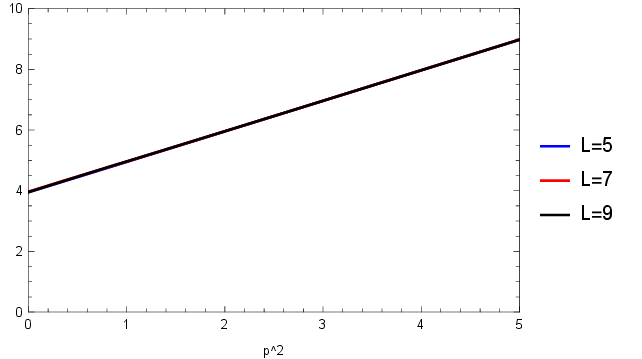}
\caption{
The smallest absolute value of eigenvalues of $C^{(g)}_{rs({\rm o})}(p)$ for $\Psi_{\rm D}$ against $\alpha^{\prime}p^2$ in the vector twist-odd sector for the ghost number $g=1$ (upper), $g=2$ (middle), and $g=3$ (lower).
\label{fig:Dsolvectorodd_Abs}
}
\end{figure}

We demonstrate the smallest absolute value of eigenvalues of the matrix $C^{(g)}_{rs({\rm e}/{\rm o})}(p)$ for various values of $\alpha^{\prime}p^2$ in Figs.~\ref{fig:Dsolscalareven_Abs},  \ref{fig:Dsolscalarodd_Abs}, \ref{fig:Dsolvectoreven_Abs}, and \ref{fig:Dsolvectorodd_Abs} for four sectors.
As in these figures, we have evaluated  $C^{(g)}_{rs({\rm e}/{\rm o})}(p)$  in the range  $-5\le \alpha^{\prime}p^2\le 0$ for the massive region and $0\le \alpha^{\prime}p^2\le 5$ for the tachyonic region. 
In our calculations, we took a difference of $\alpha^{\prime}p^2$ as $0.005$ and joined the adjacent data points for each truncation level $L$ with line segments in the figures,
where $L=4,6,8,10$ ($L=3,5,7,9$) for twist-even (twist-odd) sector.
We note that there are no vector states for $g=3$ and $L=3$ (Table~\ref{tab:Nvector}). 
In the $g=4$ and $g=5$ sectors, we have verified that $C^{(g)}_{rs({\rm e}/{\rm o})}(p)$ does not have zero eigenvalues in the above range  $-5\le \alpha^{\prime}p^2\le 5$, and thus we omit figures for them.
We observe the following from the figures for the ghost number $g=1,2,3$.

In Fig.~\ref{fig:Dsolscalareven_Abs}  for the scalar twist-even sector, zero eigenvalues of $C^{(g)}_{rs({\rm e})}(p)$ cannot be found for the tachyonic region, but there might be zero eigenvalues around $\alpha^{\prime}p^2\sim  -0$ in the $g=1$ sector (although it seems to lift with increasing truncation level) and $\alpha^{\prime}p^2\sim -2$  in the $g=3$ sector. 
In Fig.~\ref{fig:Dsolscalarodd_Abs} for the scalar twist-odd sector, zero eigenvalues of $C^{(g)}_{rs({\rm o})}(p)$ seem not to be found for the tachyonic region, where the values seem to lift with increasing truncation level in the $g=1$ and $g=2$ sectors, but there might be zero eigenvalues around $\alpha^{\prime}p^2\sim -1$ in the $g=1$ sector and $\alpha^{\prime}p^2\sim -1, -1.6$ in the $g=2$ sector.
In Fig.~\ref{fig:Dsolvectoreven_Abs} for the vector twist-even sector, zero eigenvalues of $C^{(g)}_{rs({\rm e})}(p)$ cannot be found for the tachyonic region, but there might be zero eigenvalues around 
$\alpha^{\prime}p^2\sim -2$ in the $g=1$ sector and $\alpha^{\prime}p^2\sim -0.6, -2$ in the $g=2$ sector.
In Fig.~\ref{fig:Dsolvectorodd_Abs} for the vector twist-odd sector, zero eigenvalues of $C^{(g)}_{rs({\rm o})}(p)$ might be found for $\alpha^{\prime}p^2\sim 0.6, -1.2$ in the $g=1$ sector.

\begin{figure}[htbp]
\centering
\includegraphics[width=7.5cm]{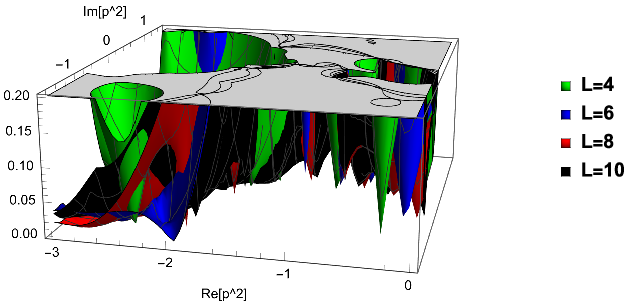}
\hfill
\includegraphics[width=7.5cm]{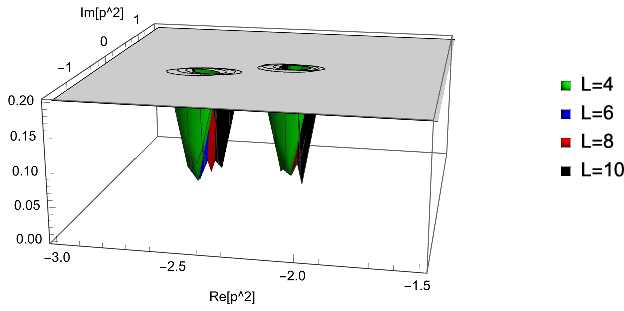}
\caption{The smallest absolute value of eigenvalues of $C^{(g)}_{rs({\rm e})}(p)$ for $\Psi_{\rm D}$ against the complex $\alpha^{\prime}p^2$ in the scalar twist-even sector for the ghost number $g=1$ (left) and $g=3$ (right).
\label{fig:Dsolscalarevencpx}
}
\end{figure}

\begin{figure}[htbp]
\centering
\includegraphics[width=7.5cm]{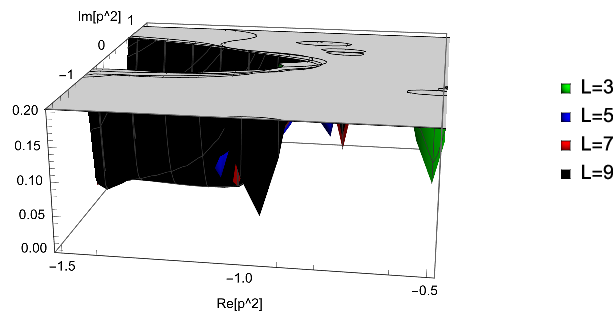}
\hfill
\includegraphics[width=7.5cm]{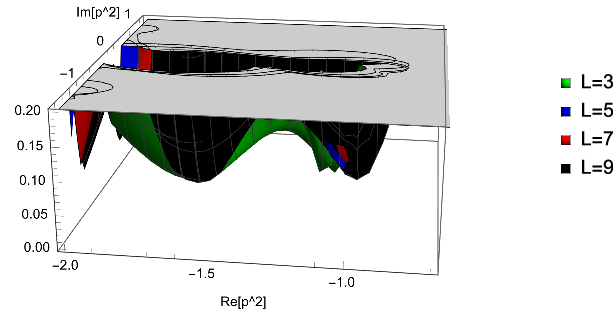}
\caption{The smallest absolute value of eigenvalues of $C^{(g)}_{rs({\rm o})}(p)$ for $\Psi_{\rm D}$ against the complex $\alpha^{\prime}p^2$ in the scalar twist-odd sector for the ghost number $g=1$ (left) and $g=2$ (right).
\label{fig:Dsolscalaroddcpx}
}
\end{figure}

\begin{figure}[htbp]
\centering
\includegraphics[width=7.5cm]{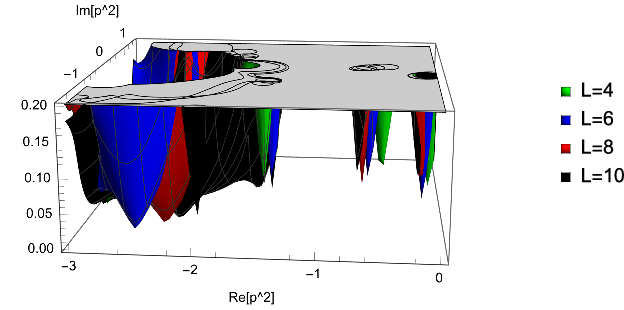}
\hfill
\includegraphics[width=7.5cm]{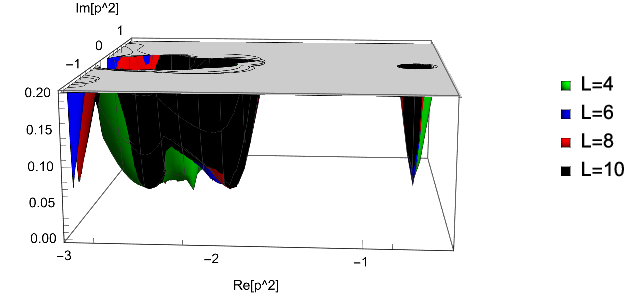}
\caption{The smallest absolute value of eigenvalues of $C^{(g)}_{rs({\rm e})}(p)$ for $\Psi_{\rm D}$ against the complex $\alpha^{\prime}p^2$ in the vector twist-even sector for the ghost number $g=1$ (left) and $g=2$ (right).
\label{fig:Dsolvectorevencpx}
}
\end{figure}

\begin{figure}[htbp]
\centering
\includegraphics[width=7.5cm]{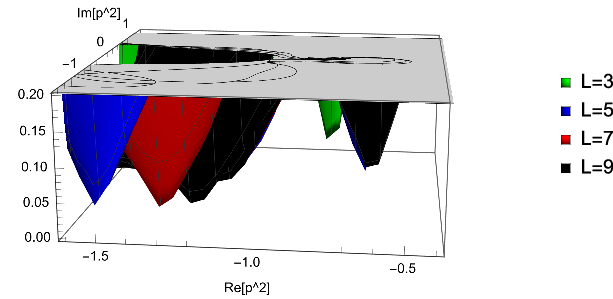}
\hfill
\includegraphics[width=7.5cm]{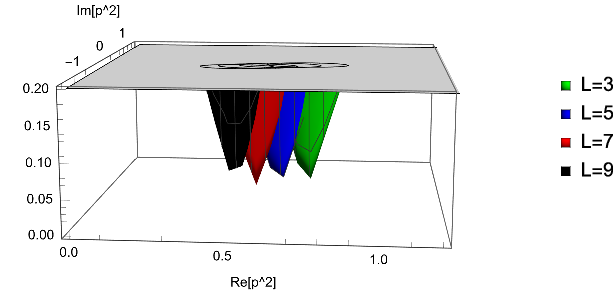}
\caption{The smallest absolute value of eigenvalues of $C^{(g)}_{rs({\rm o})}(p)$ for $\Psi_{\rm D}$ against the complex $\alpha^{\prime}p^2$ in the vector twist-odd sector for the ghost number $g=1$.
\label{fig:Dsolvectoroddcpx}
}
\end{figure}

We note that the candidates of values of $\alpha^{\prime}p^2$ for zero eigenvalues mentioned above are ambiguous compared to those for the tachyon vacuum solution $\Psi_{\rm T}$ (Appendix \ref{sec:Tresults}).
Unlike $\Psi_{\rm T}$, the ``double brane" solution $\Psi_{\rm D}$ is a complex solution at the finite truncated level, and therefore it might be necessary to include the imaginary part in $\alpha^{\prime}p^2$ to explore numerical behavior of zero eigenvalues of  $C^{(g)}_{rs({\rm e}/{\rm o})}(p)$.
Therefore, we have evaluated the eigenvalues of the matrix $C^{(g)}_{rs({\rm e}/{\rm o})}(p)$ for $\Psi_{\rm D}$ including the imaginary part of $\alpha^{\prime}p^2$ in the range $|{\rm Im}(\alpha^{\prime}p^2)|\le 1.5$. 
We regarded this complex region as massive (tachyonic) if the real part of $\alpha^{\prime}p^2$ is negative (positive).
Some results for the smallest absolute value of eigenvalues are depicted in Figs.~\ref{fig:Dsolscalarevencpx}, \ref{fig:Dsolscalaroddcpx}, \ref{fig:Dsolvectorevencpx}, and \ref{fig:Dsolvectoroddcpx}, which correspond to the characteristic regions of $\alpha^{\prime}p^2$ mentioned above.
However, these dependencies on the imaginary part seem not to show clear trends for zero eigenvalues.

Eventually, we could not find zero eigenvalues of  $C^{(g)}_{rs({\rm e}/{\rm o})}(p)$ for $\Psi_{\rm D}$ such that the corresponding mass spectrum is consistent with the theory on a double D-brane.

\section{Numerical results for the ``single brane'' solution
\label{sec:Sresult}}

The identity-based solution constructed in Ref.~\cite{Takahashi:2002ez} has a real parameter $a$, and it becomes a nontrivial solution at $a=-1/2$, which we denote as $\Psi_{\rm TT}$.
The quadratic term of the theory around $\Psi_{\rm TT}$ from Eq.~(\ref{eq:S_PsiTT}) is given by the BRST operator of the form
\begin{align}
Q_{\rm TT}=\frac{1}{2}Q_0-\frac{1}{4}(Q_2+Q_{-2})+2c_0+c_2+c_{-2}
\end{align}
instead of the conventional $Q_{\rm B}$ for the perturbative vacuum.
Here, $c_n$ is the mode of the ghost $c(z)$,  and $Q_n$ is the mode of the BRST current $j_{\rm B}(z)=cT^{\rm mat}(z)+bc\partial c(z)+\frac{3}{2}\partial^2c(z)$, which is primary.
It has been expected that this theory represents the tachyon vacuum as mentioned in Sect.~\ref{sec:Introduction}.
In the theory with $Q_{\rm TT}$, a numerical solution $\Phi_{\rm S}$ in Siegel gauge was constructed, and it has been expected that it corresponds to a single D-brane or the perturbative vacuum \cite{Kishimoto:2009nd}. 
Because this ``single brane" solution $\Phi_{\rm S}$ is a real solution to Eq.~(\ref{eq:EOMTT}) and made of twist-even SU(1,1) singlet states with zero momentum as is the case with the tachyon vacuum solution $\Psi_{\rm T}$, 
we can analyze the mass spectrum of theory around $\Phi_{\rm S}$ in a similar way to $\Psi_{\rm T}$.
Namely, in the definition of $Q^{\prime}$ from Eq.~(\ref{eq:Qprimedef}) 
we replace $Q_{\rm B}$ with $Q_{\rm TT}$ and we take $\Phi_{\rm S}$ as $\varPsi_0$, 
and then we use 
\begin{align}
\{b_0,Q_{\rm TT}\}=\frac{1}{2}L_0-\frac{1}{4}(L_2^{\rm mat}+L_2^{{\rm gh}\prime}+L_{-2}^{\rm mat}+L_{-2}^{{\rm gh}\prime})+2
\end{align}
instead of $\{b_0,Q_{\rm B}\}=L_0$ in $L_0^{\prime}=\{b_0,Q^{\prime}\}=\{b_0,Q_{\rm TT}\}+\cdots$.
Here, $L_n^{\rm mat}$ is the matter Virasoro generator, which is the mode of $T^{\rm mat}(z)$, and 
$L^{{\rm gh}\prime}_n$ is the twisted Virasoro generator with the central charge $-2$.

\begin{figure}[htbp]
\includegraphics[width=8.4cm]{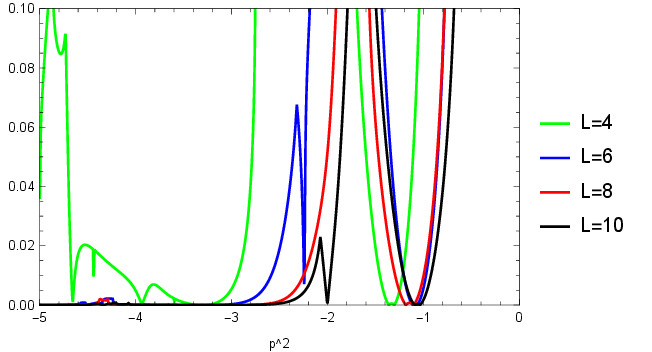}
\hfill
\includegraphics[width=8.4cm]{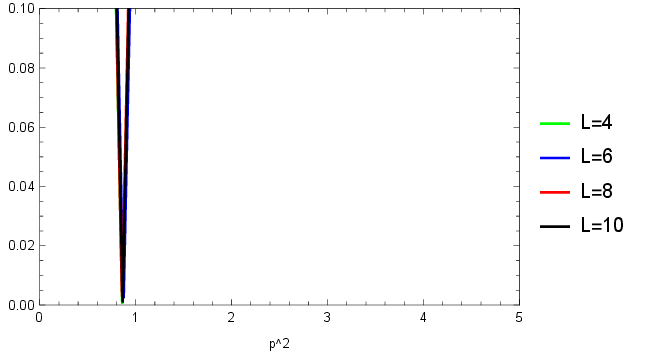}\\

\includegraphics[width=8.4cm]{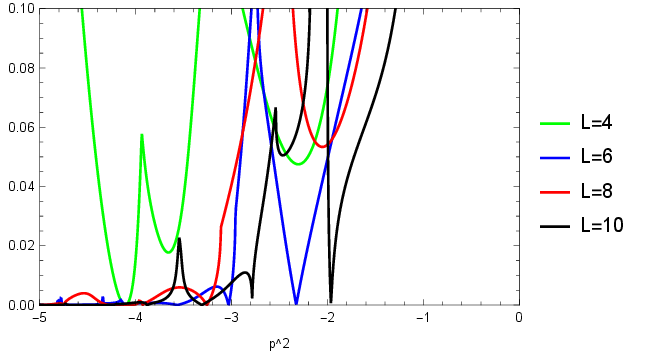}
\hfill
\includegraphics[width=8.4cm]{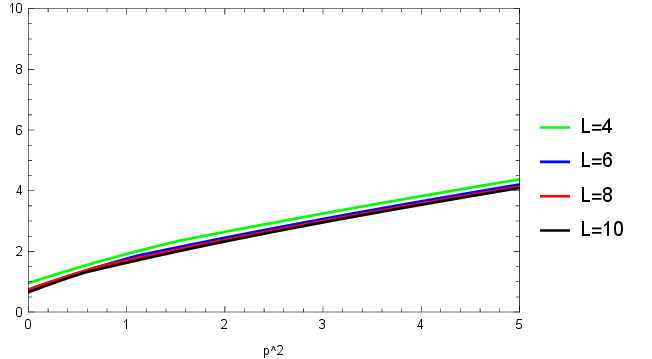}\\

\includegraphics[width=8.4cm]{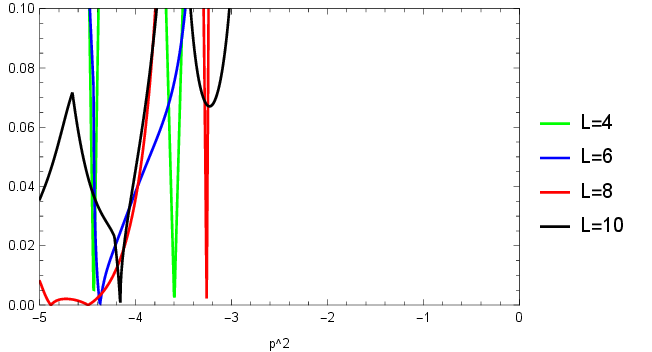}
\hfill
\includegraphics[width=8.4cm]{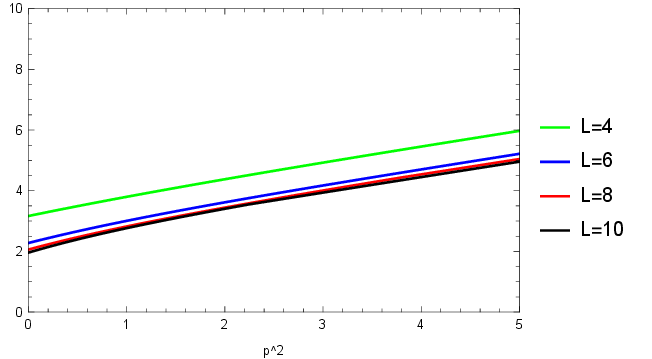}
\caption{
The smallest absolute value of eigenvalues of $C^{(g)}_{rs({\rm e})}(p)$ for $\Phi_{\rm S}$ against $\alpha^{\prime}p^2$ in the scalar twist-even sector for the ghost number $g=1$ (upper), $g=2$ (middle), and $g=3$ (lower).
\label{fig:KTSsolscalareven_Abs}
}
\end{figure}

\begin{figure}[htbp]
\includegraphics[width=8.4cm]{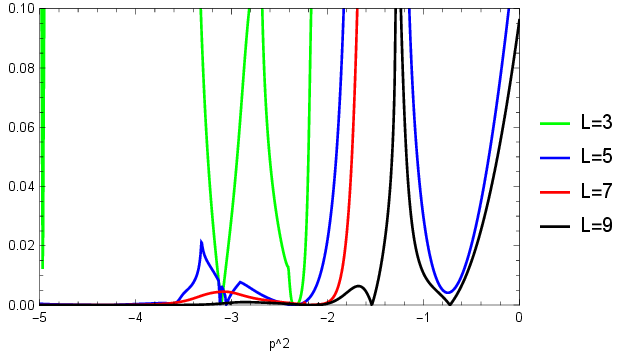}
\hfill
\includegraphics[width=8.4cm]{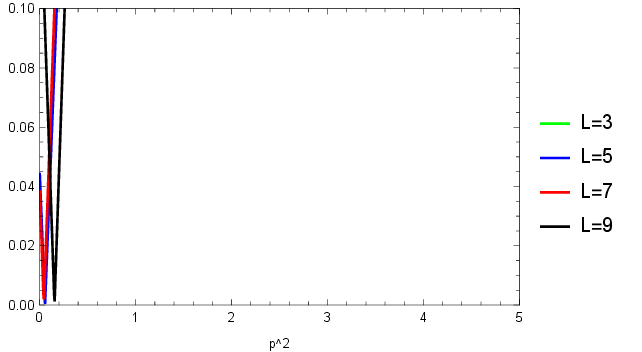}\\

\includegraphics[width=8.4cm]{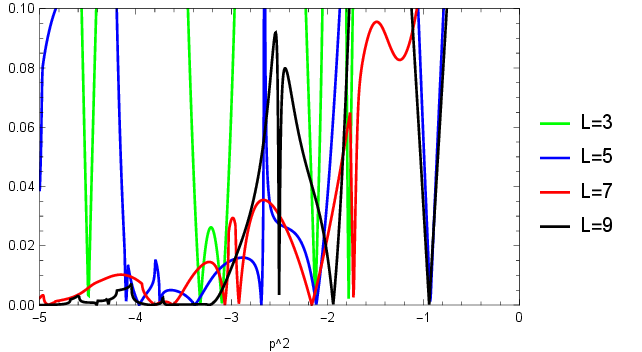}
\hfill
\includegraphics[width=8.4cm]{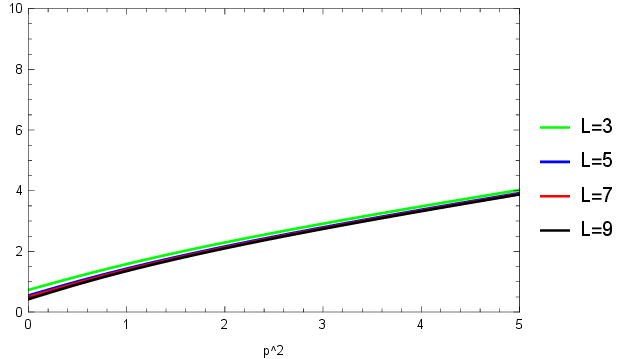}\\

\includegraphics[width=8.4cm]{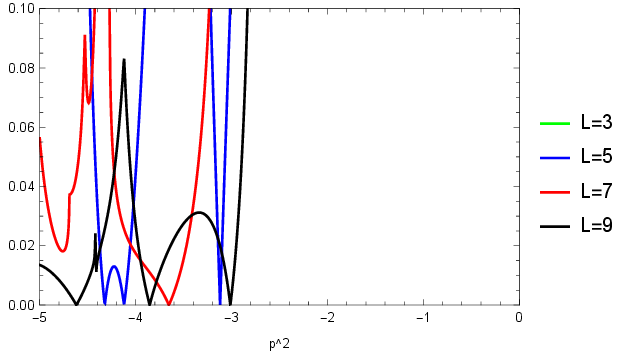}
\hfill
\includegraphics[width=8.4cm]{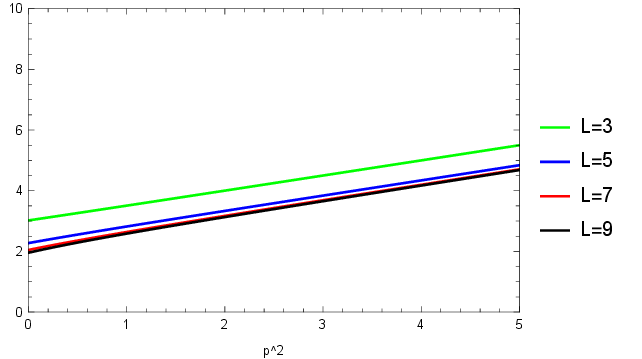}
\caption{
The smallest absolute value of eigenvalues of $C^{(g)}_{rs({\rm o})}(p)$ for $\Phi_{\rm S}$ against $\alpha^{\prime}p^2$ in the scalar twist-odd sector for the ghost number $g=1$ (upper), $g=2$ (middle), and $g=3$ (lower).
\label{fig:KTSsolscalarodd_Abs}
}
\end{figure}

\begin{figure}[htbp]
\includegraphics[width=8.4cm]{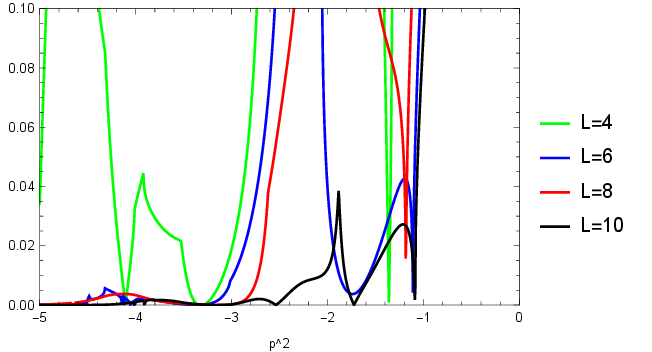}
\hfill
\includegraphics[width=8.4cm]{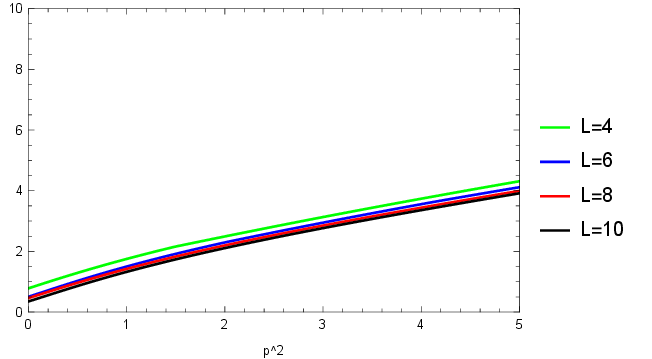}\\

\includegraphics[width=8.4cm]{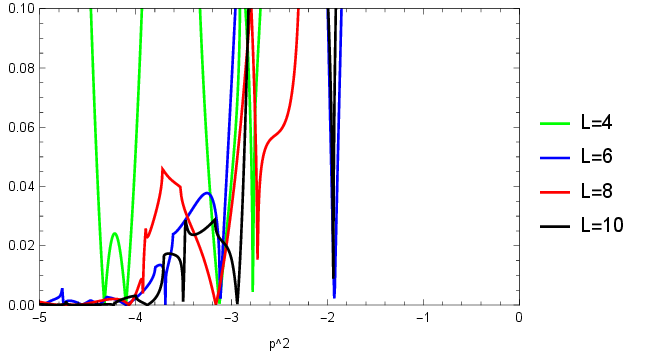}
\hfill
\includegraphics[width=8.4cm]{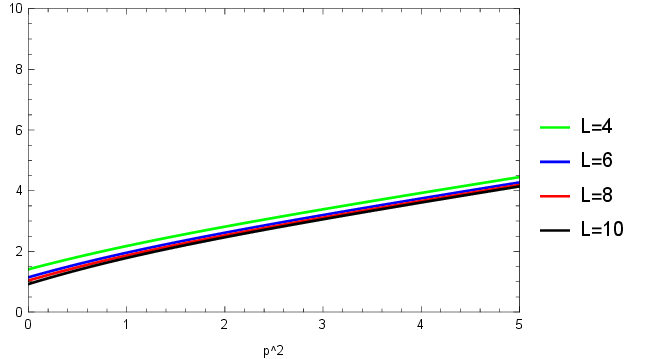}\\

\includegraphics[width=8.4cm]{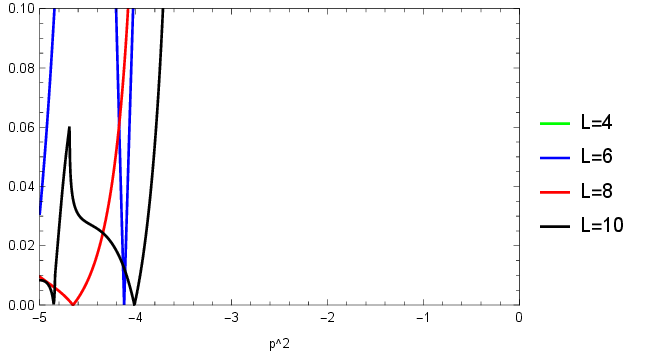}
\hfill
\includegraphics[width=8.4cm]{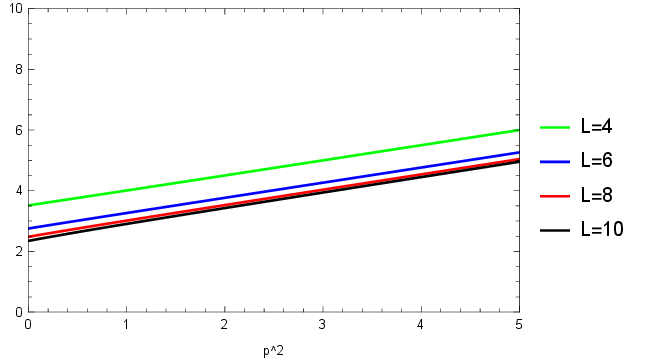}
\caption{
The smallest absolute value of eigenvalues of $C^{(g)}_{rs({\rm e})}(p)$ for $\Phi_{\rm S}$ against $\alpha^{\prime}p^2$ in the vector twist-even sector for the ghost number $g=1$ (upper), $g=2$ (middle), and $g=3$ (lower).
\label{fig:KTSsolvectoreven_Abs}
}
\end{figure}

\begin{figure}[htbp]
\includegraphics[width=8.4cm]{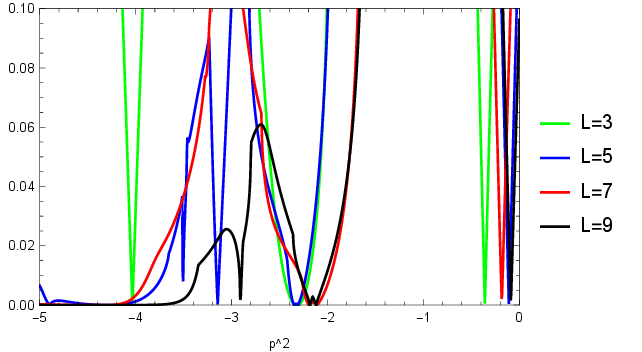}
\hfill
\includegraphics[width=8.4cm]{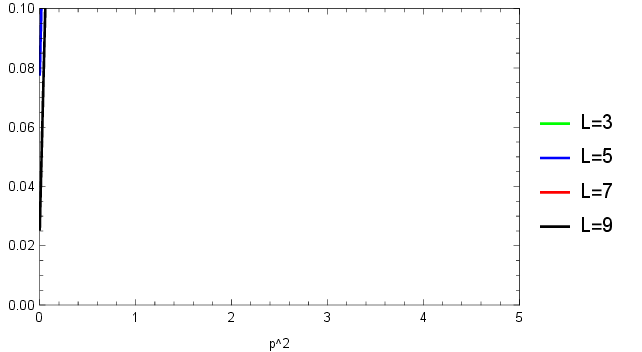}\\

\includegraphics[width=8.4cm]{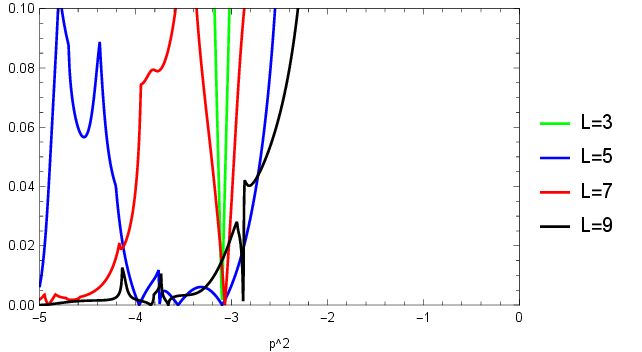}
\hfill
\includegraphics[width=8.4cm]{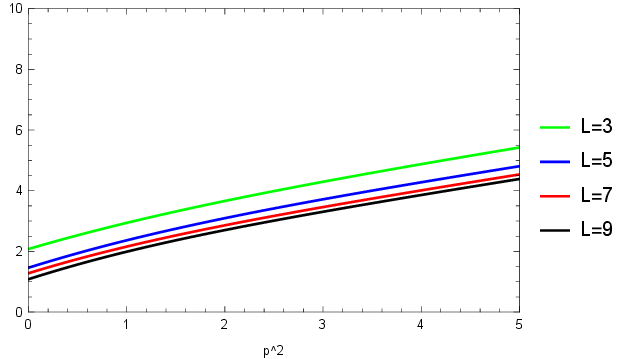}\\

\includegraphics[width=8.4cm]{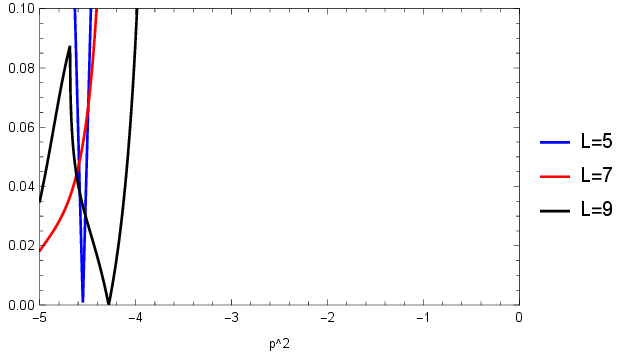}
\hfill
\includegraphics[width=8.4cm]{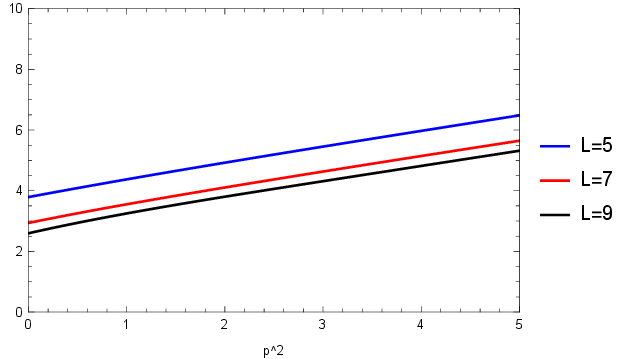}
\caption{
The smallest absolute value of eigenvalues of $C^{(g)}_{rs({\rm o})}(p)$ for $\Phi_{\rm S}$ against $\alpha^{\prime}p^2$ in the vector twist-odd sector for the ghost number $g=1$ (upper), $g=2$ (middle), and $g=3$ (lower).
\label{fig:KTSsolvectorodd_Abs}
}
\end{figure}

We show the numerical results for $\Phi_{\rm S}$ similarly to the tachyon vacuum $\Psi_{\rm T}$ in Appendix \ref{sec:Tresults}.
Firstly, as in Sect.~\ref{sec:Dresult}, we demonstrate the smallest absolute value of eigenvalues of the matrix $C^{(g)}_{rs({\rm e}/{\rm o})}(p)$ for various values of $\alpha^{\prime}p^2$ in Figs.~\ref{fig:KTSsolscalareven_Abs}, \ref{fig:KTSsolscalarodd_Abs}, \ref{fig:KTSsolvectoreven_Abs}, and \ref{fig:KTSsolvectorodd_Abs} for four sectors. 
As in these figures, we have evaluated  $C^{(g)}_{rs({\rm e}/{\rm o})}(p)$  in the range  $-5\le \alpha^{\prime}p^2\le 0$ for the massive region and $0\le \alpha^{\prime}p^2\le 5$ for the tachyonic region. 
In the cases that $g\ge 4$, we have verified that $C^{(g)}_{rs({\rm e}/{\rm o})}(p)$ does not have zero eigenvalues in the range $-5\le \alpha^{\prime}p^2\le 5$ up to the truncation level $L=10$.
We observe the following from the figures for the ghost number $g=1,2,3$.

In Fig.~\ref{fig:KTSsolscalareven_Abs} for the scalar twist-even sector, 
we can find a zero eigenvalue of the matrix $C^{(g)}_{rs({\rm e})}(p)$ around $\alpha^{\prime}p^2\sim 1$ (1t) and two zero eigenvalues around $\alpha^{\prime}p^2\sim -1$ (1a, 1b) both in the $g=1$ sector.
In Fig.~\ref{fig:KTSsolscalarodd_Abs} for the scalar twist-odd sector,
we can find a zero eigenvalue of the matrix $C^{(g)}_{rs({\rm o})}(p)$ around $\alpha^{\prime}p^2\sim +0$ (1z') in the $g=1$ sector.
In Fig.~\ref{fig:KTSsolvectoreven_Abs} for the vector twist-even sector, the existence of zero eigenvalues of the matrix $C^{(g)}_{rs({\rm e})}(p)$ seems to be ambiguous.
In Fig.~\ref{fig:KTSsolvectorodd_Abs} for the vector twist-odd sector, 
we can find a zero eigenvalue of the matrix $C^{(g)}_{rs({\rm o})}(p)$ around $\alpha^{\prime}p^2\sim -0$ (1z) and two zero eigenvalues around $\alpha^{\prime}p^2\sim -2$ (1c, 1d) both in the $g=1$ sector.
Here, (1t), (1a), (1b), (1z'), (1z), (1c), and  (1d) are labels of zero eigenvalues for the explanation below.

\begin{figure}[htbp]
\centering
\includegraphics[width=9cm]{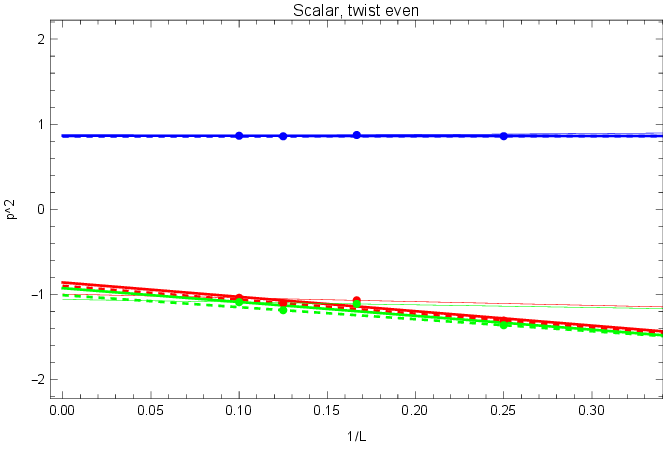}\\
\includegraphics[width=9cm]{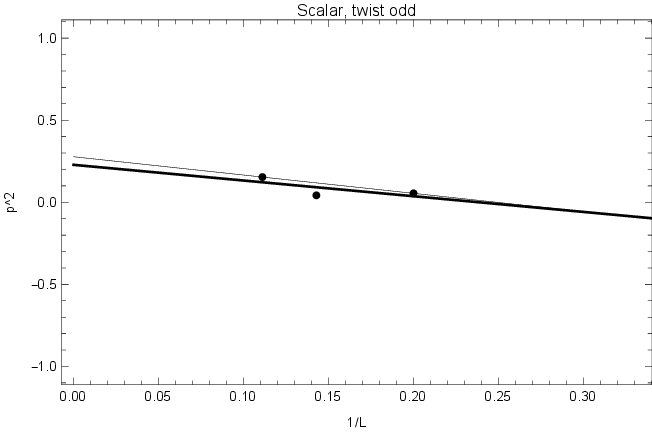}\\
\includegraphics[width=9cm]{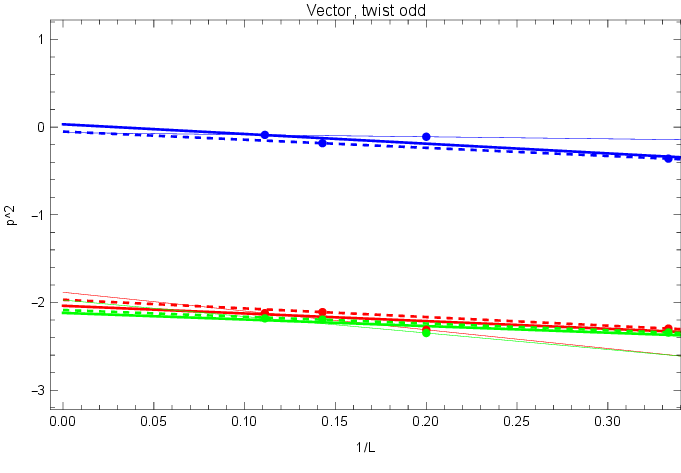}
\caption{The linear extrapolations for the values of $\alpha^{\prime}p^2$ near $0$ for zero eigenvalues against $1/L$
corresponding to Table~\ref{tab:KTSsolscalareven} (upper), Table~\ref{tab:KTSsolscalarodd} (middle), and Table~\ref{tab:KTSsolvectorodd} (lower).
\label{fig:KTSsol_fit}
}

\end{figure}

We precisely computed the eigenvalues of $C^{(g)}_{rs({\rm e}/{\rm o})}(p)$ around the values of $\alpha^{\prime} p^2$ mentioned above and found the precise values of $\alpha^{\prime}p^2$ which give zero eigenvalues.
They are listed in Tables~\ref{tab:KTSsolscalareven}, \ref{tab:KTSsolscalarodd}, and \ref{tab:KTSsolvectorodd}.
In these tables, $L=2\infty$ or $L=2\infty+1$ denotes the extrapolation with a fit by $\mathcal{C}_1/L+\mathcal{C}_0$, where $\mathcal{C}_0$ and $\mathcal{C}_1$ are constants, using the data for $L=2n$ or $L=2n+1$ ($n$ is an integer).
We note that some quantities for $\Phi_{\rm S}$, which is a twist-even solution in Siegel gauge constructed by the level truncation approximation in the theory around $\Psi_{\rm TT}$, show good numerical behavior by fitting with the interval level $4$ as in Refs.~\cite{Kishimoto:2011zza, Kishimoto:2020vfg}.
Therefore, in Table~\ref{tab:KTSsolscalareven}, we also write down the extrapolations using data for $L=4n$ and  $L=4n+2$ and denote them as $4\infty$ and $4\infty+2$, respectively.
Similarly, in Tables \ref{tab:KTSsolscalarodd} and \ref{tab:KTSsolvectorodd},
 we also write down the extrapolations using the data for $L=4n-1$ and $L=4n+1$ and denote them as $4\infty-1$ and $4\infty+1$, respectively.
However, we have few data in these cases of the truncated level up to $L=10$ or $L=9$,
 and hence it is expected to have more data for higher levels to get more reliability.
In Fig.~\ref{fig:KTSsol_fit} (upper), blue, red, and green lines are the fit lines for (1t), (1a), and (1b) in Table~\ref{tab:KTSsolscalareven}, respectively (solid: $L=2n$, dashed: $L=4n$ and thin: $L=4n+2$).
In Fig.~\ref{fig:KTSsol_fit} (middle and lower), black, blue, red, and green lines are the fit lines for (1z'), (1z), (1c), and (1d) in Tables~\ref{tab:KTSsolscalarodd} and \ref{tab:KTSsolvectorodd}, respectively (solid: $L=2n+1$, dashed: $L=4n-1$ and thin: $L=4n+1$).

\begin{table}[htbp]
\caption{The values of $\alpha^{\prime}p^2$ corresponding to a zero eigenvalue of $C^{(g=1)}_{rs({\rm e})}(p)$ in the scalar twist-even sector with the ghost number $g=1$.
\label{tab:KTSsolscalareven}}
\begin{center}
\begin{tabular}{c|| l | l | l | l | l}
\hline
label&$L=4$&$L=6$&$L=8$&$L=10$&$L=2\infty,~4\infty,~4\infty+2$\\
\hline\hline
(1t)&$0.859725$&$0.873436$&$0.858369$&$0.864498$&$0.86592$,
~$0.857013$,
~$0.851092$\\
 \hline
(1a)&$-1.30561$&$-1.06781$&$-1.10165$&$-1.03842$&$-0.85774$,
~$-0.897704$,
~$-0.994344$\\
 \hline
 (1b)&$-1.3596$&$-1.10969$&$-1.18316$&$-1.08858$&$-0.924749$,
 ~$-1.00672$,
 ~$-1.05693$\\
\hline
      \end{tabular}
    \end{center}
\end{table}

\begin{table}[htbp]
\caption{The values of $\alpha^{\prime}p^2$ corresponding to a zero eigenvalue of $C^{(g=1)}_{rs({\rm o})}(p)$ in the scalar twist-odd sector with the ghost number $g=1$.
\label{tab:KTSsolscalarodd}}
\begin{center}
\begin{tabular}{c|| l | l | l | l l |}
\hline
label&$L=5$&$L=7$&$L=9$&$L=2\infty+1,~4\infty+1$\\
\hline\hline
 (1z')&$0.0545303$&$0.0423039$&$0.153635$&$0.228232$,
 ~$0.277515$\\
\hline
      \end{tabular}
    \end{center}
\end{table}

\begin{table}[htbp]
\caption{The values of $\alpha^{\prime}p^2$ corresponding to a zero eigenvalue of $C^{(g=1)}_{rs({\rm o})}(p)$ in the vector twist-odd sector with the ghost number $g=1$.
\label{tab:KTSsolvectorodd}}
\begin{center}
\begin{tabular}{c|| l | l | l | l | l}
\hline
label&$L=3$&$L=5$&$L=7$&$L=9$&$L=2\infty+1,~4\infty-1,~4\infty+1$\\
\hline\hline
 (1z)&$-0.359597$&$-0.109688$&$-0.183161$&$-0.0885834$&$0.032367$,
 ~$-0.0508333$,
 ~$-0.0622024$\\
 \hline
 (1c)&$-2.29722$&$-2.31091$&$-2.10953$&$-2.12089$&$-2.03744$,
 ~$-1.96876$,
 ~$-1.88336$\\
 \hline
 (1d)&$-2.34204$&$-2.34789$&$-2.19564$&$-2.18084$&$-2.11858$,
 ~$-2.08584$,
 ~$-1.97203$\\
\hline
      \end{tabular}
    \end{center}
\end{table}

We have numerically confirmed that all states constructed from eigenvectors corresponding to zero eigenvalues, (1t), (1a), (1b), (1z'), (1z), (1c), and  (1d), are SU(1,1) singlet.

As seen from the above, we have numerically found the following mass spectrum made of SU(1,1)-singlet states in the ghost number $g=1$ sector in the theory around $\Phi_{\rm S}$:
one scalar twist-even state with $\alpha^{\prime}m^2\simeq -1$ (1t);
two scalar twist-even states with $\alpha^{\prime}m^2\simeq 1$ (1a, 1b);
one scalar twist-odd state with $\alpha^{\prime}m^2\simeq 0$ (1z');
one vector twist-odd state with $\alpha^{\prime}m^2\simeq 0$ (1z);
two vector twist-odd states with $\alpha^{\prime}m^2\simeq 2$ (1c, 1d).
In particular, (1t) corresponds to the tachyon state on a D-brane.
Furthermore, (1z') and (1z) have $1+25=26$ components and correspond to the massless state on a D-brane.
Therefore, these results are consistent with the interpretation that $\Phi_{\rm S}$ represents a single brane solution in the theory around the tachyon vacuum given by $\Psi_{\rm TT}$.
However, we could not find zero eigenvalues corresponding to massless scalar states in the ghost number $g=2$ and $g=0$ sectors, which exist in the conventional theory with $Q_{\rm B}$. 
To confirm the consistency for massive states in the theory around $\Phi_{\rm S}$, tensor states should be included to calculate $C^{(g)}_{rs({\rm e}/{\rm o})}(p)$ in addition to scalar and vector states.

\section{Concluding remarks
\label{sec:conclusion}}

In this paper, we have investigated the kinetic term in the gauge-fixed theory around the ``double brane" solution $\Psi_{\rm D}$ and the ``single brane" solution $\Phi_{\rm S}$ in Siegel gauge using the numerical method in Refs.~\cite{Giusto:2003wc, Imbimbo:2006tz} for the tachyon vacuum solution $\Psi_{\rm T}$ in Siegel gauge.
$\Psi_{\rm D}$ is a numerical solution to the theory around the perturbative vacuum and 
$\Phi_{\rm S}$ is a numerical solution to the theory around the identity-based solution $\Psi_{\rm TT}$ for the tachyon vacuum.
We have evaluated the eigenvalues of the matrix $C^{(g)}_{rs({\rm e}/{\rm o})}(p)$ of the kinetic term for the scalar and vector states up to the truncation level $L=10$.

In the theory around $\Psi_{\rm D}$, we did not have a definite result because the existence of zero eigenvalues of $C^{(g)}_{rs({\rm e}/{\rm o})}(p)$ was ambiguous. Therefore, we could not obtain any evidence that $\Psi_{\rm D}$ can be interpreted as a solution corresponding to a double brane.

On the other hand, in the theory around $\Phi_{\rm S}$, we have found a tachyon state with $m^2\simeq -1/\alpha^{\prime}$ and a vector state of $1+25=26$ components with $m^2\simeq 0$ in the ghost number $g=1$ sector, and these are consistent with the mass spectrum of the theory on a D-brane.
This result supports the previous interpretation that $\Phi_{\rm S}$ represents the perturbative vacuum (a single brane).
However, we should note that there exist massless scalar states in the ghost number $g=2$ and $g=0$ sectors in the conventional theory of $Q_{\rm B}$, which we could not find.
They are necessary to get the correct physical degrees of freedom such as $26-2=24$.
In this sense, the massless spectrum which we have found in this paper is still incomplete.

In our computation of the kinetic term around numerical solutions, we have used {\sl Mathematica} and it seems to be difficult to struggle with higher truncation levels such as $L>10$ in the same way, although we have numerical data for solutions $\Psi_{\rm T}$, $\Psi_{\rm D}$, and $\Phi_{\rm S}$ themselves for higher levels in our previous work \cite{Kishimoto:2020vfg}.
If we perform computations for higher truncation levels, we will get further data for the extrapolations of the mass spectrum in the large-$L$ limit, but we have to develop more efficient methods and code for numerical calculations.
If we can extract some physical meaning for the ``double brane" solution from higher-level data, it may be interesting to apply the method to the ``ghost brane" solution constructed in Ref.~\cite{Kudrna:2018mxa}.

Furthermore, we should include tensor states with higher rank in addition to scalar and vector states to investigate more massive states in the theory around numerical solutions, and computational developments for higher truncation levels are indispensable for such direction.

It is an important future problem to confirm the BRST invariance of the states corresponding to the zero eigenvalues of  $C^{(g)}_{rs({\rm e}/{\rm o})}(p)$ numerically.
In Refs.~\cite{Giusto:2003wc, Imbimbo:2006tz}, the BRST invariance of them around $\Psi_{\rm T}$ was investigated from the consistency of the relative and absolute cohomologies in the context of the semi-infinite exact sequences.
It seems that we do not have the established numerical criteria on the BRST invariance \cite{Hata:2000bj}, or the out-of-Siegel-gauge equation, even for solutions themselves in Siegel gauge, and hence it is preferable to develop appropriate numerical methods to evaluate the BRST invariance around the solutions directly.

\section*{Acknowledgments}
We would like to express our gratitude to T.~Takahashi for valuable discussions.
The author thanks the Yukawa Institute for Theoretical Physics at Kyoto University.
Discussions during the YITP workshop YITP-W-23-07 on ``Strings and Fields 2023'' were useful in completing this work.
This work was supported in part by JSPS KAKENHI Grant Numbers JP20K03933, JP20K03972. The numerical calculations were partly carried out on sushiki at YITP in Kyoto University.

\appendix

\section{Numerical results for the tachyon vacuum solution
\label{sec:Tresults}
}

Although numerical computations for the tachyon vacuum solution $\Psi_{\rm T}$ in Siegel gauge have been performed in Refs.~\cite{Giusto:2003wc, Imbimbo:2006tz}, we list the results up to the truncation level $L=10$ with the method in Sect.~\ref{sec:Method} for comparison.

\begin{figure}[htbp]
\centering
\includegraphics[height=4.5cm]{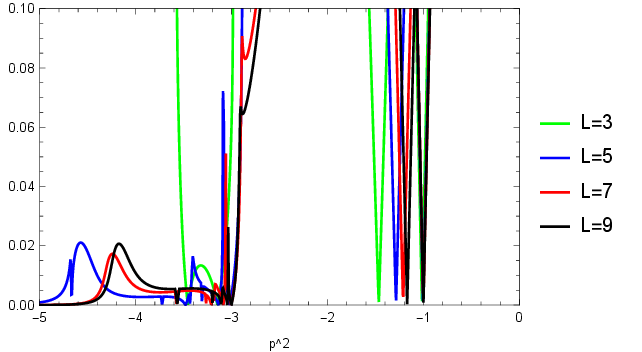}
\hfill
\includegraphics[height=4.5cm]{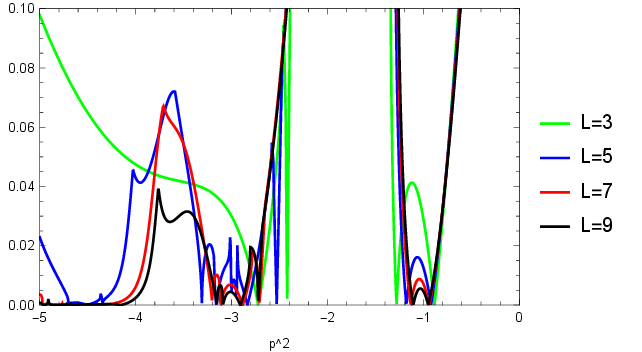}\\

\vspace{1em}

\includegraphics[height=4.5cm]{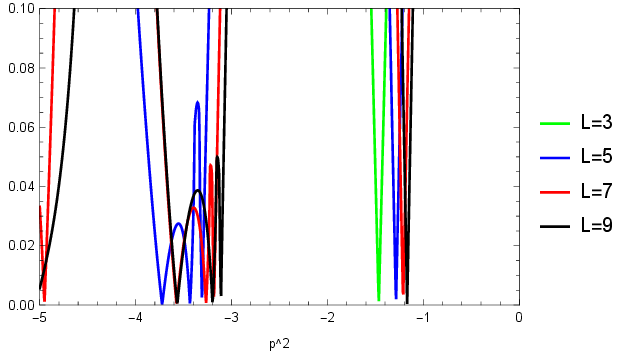}
\hfill
\raisebox{-1.5em}[4.5cm][0.5em]{
\includegraphics[height=4.7cm]{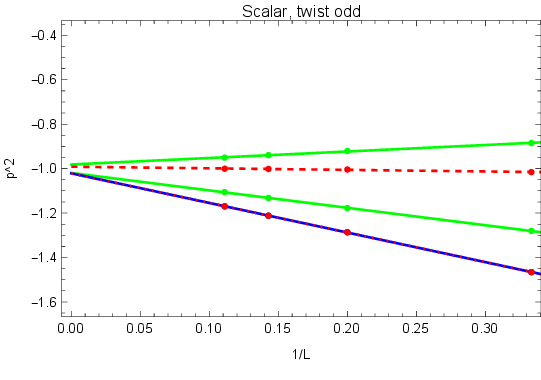}}\quad{~}

\caption{The smallest absolute value of eigenvalues of $C^{(g)}_{rs({\rm o})}(p)$ for $\Psi_{\rm T}$ against $\alpha^{\prime}p^2$ in the scalar twist-odd sector for the ghost number $g=1$ (upper left), $g=2$ (upper right), and $g=3$ (lower left).
The lower right figure denotes the linear extrapolations for the values of $\alpha^{\prime}p^2$ near $-1$ for zero eigenvalues against $1/L$. (The red broken lines correspond to the states in the $g=1$ sector, where the lower line overlaps with the blue line which corresponds to the states in the $g=3$ and $g=-1$ sectors.
The green lines correspond to the states in the $g=2$ and $g=0$ sectors.)
\label{fig:Tsolscaodd}}
\end{figure}

We list the results for the scalar twist-odd sector in Fig.~\ref{fig:Tsolscaodd}.
Three figures show the smallest absolute value of eigenvalues of the matrix $C^{(g)}_{rs({\rm o})}(p)$ for various values of $\alpha^{\prime}p^2$ with $L=3,5,7,9$ in the ghost number $g=1,2,3$ sector.
We have observed that it becomes zero near $\alpha^{\prime}p^2=-1$ and evaluated the precise values of $\alpha^{\prime}p^2$ for the zero eigenvalues. We plotted them against $1/L$ with the linear extrapolations for $L\to\infty$ in Fig.~\ref{fig:Tsolscaodd} (lower right).

\begin{figure}[htbp]
\centering
\includegraphics[height=4.5cm]{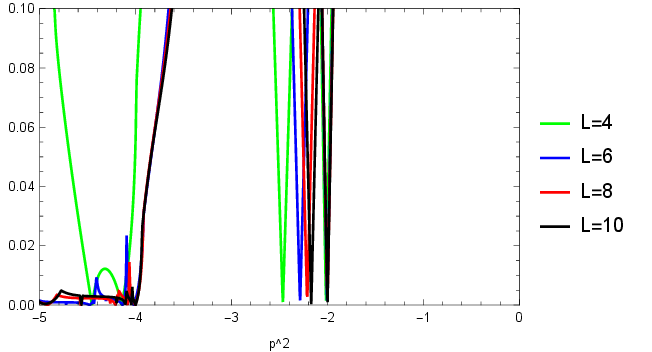}
\hfill
\includegraphics[height=4.5cm]{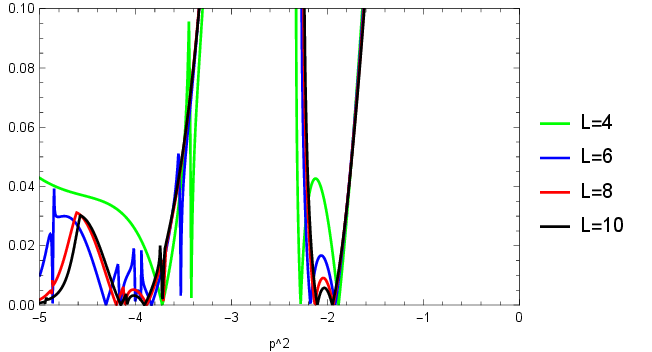}\\

\vspace{1em}

\includegraphics[height=4.5cm]{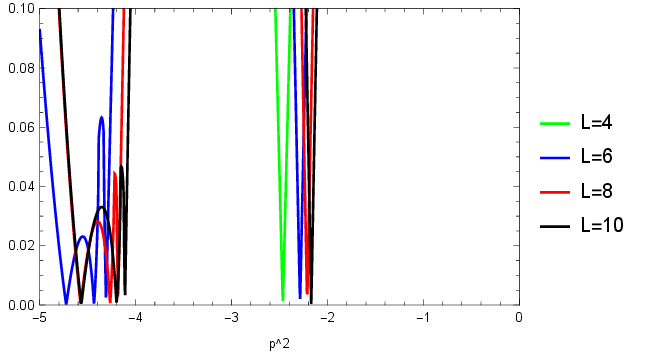}
\hfill
\raisebox{-1.5em}[4.5cm][0.5em]{
\includegraphics[height=4.7cm]{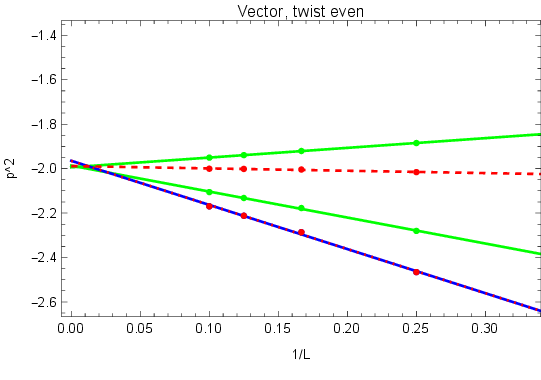}}\quad{~}

\caption{The smallest absolute value of eigenvalues of $C^{(g)}_{rs({\rm e})}(p)$ for $\Psi_{\rm T}$ against $\alpha^{\prime}p^2$ in the vector twist-even sector for the ghost number $g=1$ (upper left), $g=2$ (upper right), and $g=3$ (lower left).
The lower right figure denotes the linear extrapolations for the values of $\alpha^{\prime}p^2$ near $-2$ for zero eigenvalues against $1/L$. (The red broken lines correspond to the states in the $g=1$ sector, where the lower line overlaps with the blue line which correspond to the states in the $g=3$ and $g=-1$ sectors.
The green lines correspond to the states in the $g=2$ and $g=0$ sectors.)
\label{fig:Tsolveceven}}
\end{figure}

\begin{figure}[htbp]
\centering
\includegraphics[height=4.5cm]{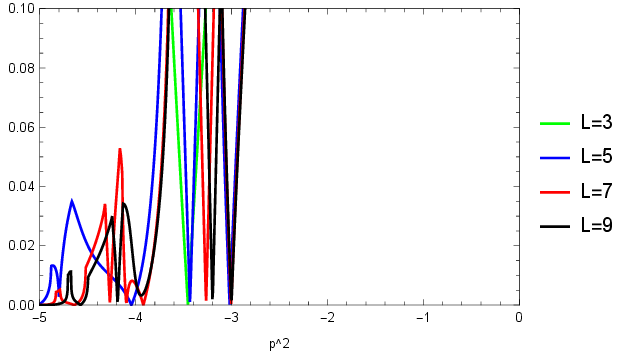}
\hfill
\includegraphics[height=4.5cm]{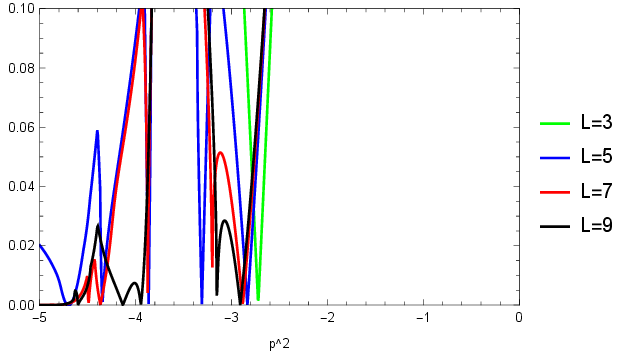}\\

\vspace{1em}

\includegraphics[height=4.5cm]{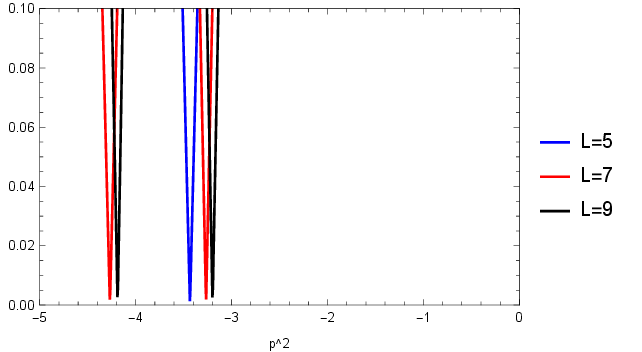}
\hfill
\raisebox{-1.5em}[4.5cm][0.5em]{
\includegraphics[height=4.7cm]{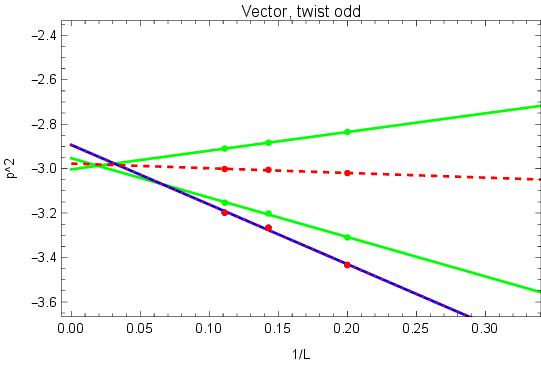}}\quad{~}
\caption{The smallest absolute value of eigenvalues of $C^{(g)}_{rs({\rm o})}(p)$ for $\Psi_{\rm T}$ against $\alpha^{\prime}p^2$ in the vector twist-odd sector for the ghost number $g=1$ (upper left), $g=2$ (upper right), and $g=3$ (lower left).
The lower right figure denotes the linear extrapolations for the values of $\alpha^{\prime}p^2$ near $-3$ for zero eigenvalues against $1/L$. (The red broken lines correspond to the states in the $g=1$ sector, where the lower line overlaps with the blue line which corresponds to the states in the $g=3$ and $g=-1$ sectors.
The green lines correspond to the states in the $g=2$ and $g=0$ sectors.)
\label{fig:Tsolvecodd}}
\end{figure}

Similarly, we list the results for the vector twist-even and vector twist-odd sectors in Fig.~\ref{fig:Tsolveceven}  and Fig.~\ref{fig:Tsolvecodd}, which show the smallest absolute value of eigenvalues of the matrix $C^{(g)}_{rs({\rm e}/{\rm o})}(p)$ for $L=4,6,8,10$ or $L=3,5,7,9$  in the ghost number $g=1,2,3$ sector.
In the vector twist-even (twist-odd) sector, it becomes zero near  $\alpha^{\prime}p^2=-2$ ($\alpha^{\prime}p^2=-3$) and we computed the precise values of $\alpha^{\prime}p^2$ for these zero eigenvalues. We plotted them against $1/L$ with the linear extrapolations for $L\to\infty$  in Fig.~\ref{fig:Tsolveceven} (lower right) and Fig.~\ref{fig:Tsolvecodd} (lower right).

After identifying the eigenstates for zero eigenvalues in each sector, we can find SU(1,1)-singlet, doublet, and triplet states with mass $\alpha^{\prime}m^2\simeq 1$ (scalar twist-odd sector), $\alpha^{\prime}m^2\simeq 2$ (vector twist-even sector) and  $\alpha^{\prime}m^2\simeq 3$ (vector twist-odd sector) \cite{Imbimbo:2006tz}, where the generators of SU(1,1) is given by
\begin{align}
&J_3=\frac{1}{2}\sum_{n=1}^{\infty}(c_{-n}b_n-b_{-n}c_n),
&&J_+=\sum_{n=1}^{\infty}nc_{-n}c_n,
&&J_-=\sum_{n=1}^{\infty}\dfrac{1}{n}b_{-n}b_n.
\end{align}

As for the scalar twist-even sector, we cannot find zero eigenvalues of $C^{(g)}_{rs({\rm e})}(p)$ such as those in the other three sectors mentioned above.
In the tachyonic region, we have evaluated the matrix $C^{(g)}_{rs({\rm e}/{\rm o})}(p)$ for $0\le \alpha^{\prime}p^2\le 5$ in four sectors (scalar twist-even/odd and vector twist-even/odd) and found no zero eigenvalues.

\bibliographystyle{utphys}
\bibliography{referencev3}

\end{document}